TECHNICAL UNIVERISTY OF ŁÓDŹ

FACULTY OF ELECTRICAL, ELECTRONIC, COMPUTER AND
CONTROL ENGINEERING

COMPUTER ENGINEERING DEPARTMENT

MASTER'S THESIS

# WEB GRAPH COMPRESSION
# WITH FAST ACCESS

## KOMPRESJA GRAFU WEBOWEGO
## Z SZYBKIM DOSTĘPEM

**Filip PROBORSZCZ**

**170421**

Supervisor:

Szymon GRABOWSKI, PhD

Additional advisor:

Wojciech BIENIECKI, PhD Eng

Łódź, May 2012

# TABLE OF CONTENTS

# ABSTRACT

In recent years studying the content of the World Wide Web became a very important yet rather difficult task. Fast resource searching engines need to explore the Internet and index its whole structure while still being profitable i.e. maintaining low machine's resource consumption that implies lower costs. In order to achieve this task a new structure was introduced, a *web graph*. Since the Internet consists of billions of pages it is rather hard to enclose it in RAM in order to process it fast. Therefore there is a need for a compression technique that would allow a web graph representation to be put into the memory while maintaining random access time competitive to the time needed to access uncompressed web graph on a hard drive.

There are already available techniques that accomplish this task, but there is still room for improvements and this thesis attempts to prove it. It includes a comparison of two methods contained in state of art of this field (*BV* and $k^2 partitioned$) to two already implemented algorithms (rewritten, however, in C++ programming language to maximize speed and resource management efficiency), which are LM and 2D, and introduces the new variant of the latter one, called 2D stripes.

This thesis serves as well as a proof of concept. The final considerations show positive and negative aspects of all presented methods, expose the feasibility of the new variant as well as indicate future direction for development.

# STRESZCZENIE

Na przestrzeni ostatnich lat badanie sieci WWW stało się zarówno istotnym, jak i dość trudnym zadaniem. Szybkie wyszukiwarki zasobów sieciowych przeszukują Internet i potrzebują zindeksować jego strukturę jednocześnie pozwalając na uzyskanie niskich kosztów, co oznacza, że powinny zużywać jak najmniej zasobów komputerowych. Aby podołać temu zadaniu, została wprowadzona nowa struktura – graf webowy. Z powodu istnienia miliardów stron w Internecie trudno jest zmieścić całą informację w pamięci RAM komputera tak, aby ją szybko móc przetwarzać. Stąd istnieje zapotrzebowanie na technikę kompresji, która umożliwiałaby pomieszczenie całego grafu webowego w pamięci komputera jednocześnie zapewniając szybkość odczytu informacji na poziomie wyższym niż prędkość odczytywania nieskompresowanej reprezentacji z dysku twardego.

Obecnie istnieją metody, które pozwalają na spełnienie tego zadania, jednak nadal da się je polepszyć i ta praca postara się to udowodnić. Zawiera ona porównanie dwóch popularnych technik (*BV* i $k^2$*partitioned*) z pozostałymi dwoma już zaimplementowanymi algorytmami (jednak przepisanymi w języku programowania C++ w celu zmaksymalizowania prędkości oraz wydajności zarządzania zasobami), *LM* oraz *2D*, z czego zaproponowany został nowy wariant ostatniej metody nazwany *2D stripes*.

Przedstawiona praca jest również dowodem koncepcji przedstawionego wariantu metody. Ostateczna dyskusja ukazuje pozytywne oraz negatywne aspekty wszystkich przedstawionych metod, pokazuje wykonalność zaproponowanego wariantu oraz wskazuje kierunek jego rozwoju.

# 1. GOAL AND SCOPE OF THE THESIS

The main goal of this thesis is to elaborate on techniques of compressing web graphs that allow fast access to the successor list of a node and to show positive and negative aspects of proposed methods and other known ones as well as their practical applications. In this work three methods were evaluated and tested: LM (for list merge), 2D-stripes and 2D-nostripes. They were then compared to BV [1], the most popular compressed web graph representation in the market, and $k^2$partitioned [2], which is also very promising.

This thesis brings the following contributions:

1. Investigation of currently available tools and algorithms for compressing web graphs,

2. Attempt to translate LM method, which is currently available in Java, into C++ in order to check its efficiency among other C++ solutions,

3. Elaboration of demanded features of the new solution,

4. Analysis of possible advantages and drawbacks of each idea and choosing most suitable algorithms for testing,

5. The implementation of 2D and the new variant: 2D stripes,

6. Comparative experimental results and discussion.

The ideas presented in this work are complete, which means that they were not created and intended for the given test set only, but can be used with any possible web graph. They are not limited to any specific demand, but are highly scalable. The main goal of creating them was to be able to fit the whole graph into the Random Access Memory (RAM) of a computer, thus the aim was to compress data as much as possible to obtain sensible size while maintaining quite low data extraction latency.

The solutions presented are written in C++ for its higher level of managing computer resources compared to e.g. Java or Python as well as its memory and speed efficiency. The code may be added to third party solution in order to perform its tasks according to the specific demand. Technologies and libraries used will be covered in the chapter describing implementation. All datasets used for testing purposes were provided by *Laboratory for Web Algorithmics*[1] (*LAW*) at *Computer Science Department* (*DSI*) of *University of Milan*[2].

The thesis is divided into 10 chapters:

➢ Chapter 1 contains the aim of the work and its structure.

➢ Chapter 2 provides definitions and theoretical background.

➢ Chapter 3 depicts the problem and presents some available solutions.

➢ Chapter 4 introduces new ideas and some theoretical discussion.

➢ Chapter 5 reveals the implementation details.

➢ Chapter 6 describes the test procedure, shows the results and discusses them.

➢ Chapter 7 is a short summary of the thesis.

➢ Chapter 8 contains the list of references used.

➢ Chapter 9 is a list of attachments.

➢ Chapter 10 shows all the acronyms used in the text.

---

[1] http://law.dsi.unimi.it/datasets.php [May, 2012]
[2] http://www.unimi.it/ [May, 2012]

# 2. INTRODUCTION

In recent years studying the content of the World Wide Web became a very important yet rather difficult task. Fast resource searching engines need to explore the Internet and index its whole structure while still being profitable i.e. maintaining low machine's resource consumption that implies lower costs. In order to achieve this task a new structure was introduced, a *web graph*.

In the field of World Wide Web analysis the term *web graph* (also called *link graph*) has a different meaning than its mathematical one. It is "an abstract data type which represents connections between two or more web documents in the Internet [2, p. 10]" rather than „a prism graph $Y_{n+1,3}$ with the edges of the outer cycle removed"[3]. That means that a web graph is constructed by web documents which are vertices and hyperlinks being edges. Documents can be described by its *Uniform Resource Locators* (*URL*), but it is not obligatory (then the vertices are labelled with unique identifiers, for example integers). The graph is usually constructed by a web crawler.

It should be noted that a web graph consists of web documents, which means not only static pages or files, but all information resources[4] that have their *Uniform Resource Identifiers* (*URI*). This signifies that dynamically created pages can also be nodes in such a graph, which increases the risk of creating highly connected graph subdivisions that represent more of social connections in a collaboration network (such as *Facebook*, *Twitter* or *LiveJournal*) created by the users than the structure of the actual web implied as a set of documents generated by independent authorities. In this thesis the latter depiction should be used as the tests were conducted using such type of web graphs.

In a mathematical sense a web graph is a directed unweighted graph. It can be cyclic (for example when a sub-webpage links to the main page creating a cycle) and due to high probability of constructing $K_5$ or $K_{3,3}$[5] it is rarely planar. By *outdegree* of a vertex $u$ we denote the number of edges incident to $u$ directed at another vertex $v$ that is called

---

[3] http://mathworld.wolfram.com/WebGraph.html [May, 2012]
[4] http://www.w3.org/TR/2004/REC-webarch-20041215/ [May, 2012]
[5] $K_5$ is a complete graph on 5 vertices; $K_{3,3}$ is a complete bipartite on six vertices, three of which connect to each of the other three

its *successor*. The *indegree* of a vertex *u* is a number of edges incident to *u* directing from another vertex *v* that is called its *predecessor*. Vertices in web graphs are also called nodes.

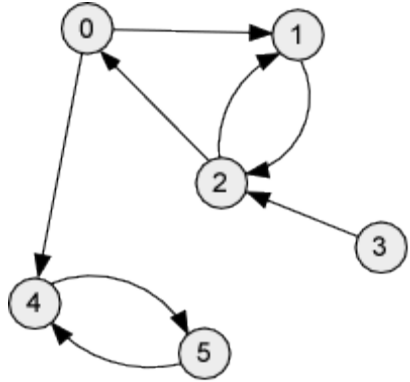

*Figure 1. An example of a graphical representation of a simple web graph*[6]

The graph can be represented in a traditional graphical way, as seen in figure 1, though it is quite impractical to do so because of unimaginably huge amount of edges and nodes. The Internet has been reported to contain around 45 billion pages.[7] It is clear that such vast amount of data must be processed by a machine rather than by human.

One of the simplest yet quite commonly used techniques of storing information about a graph is an *adjacency matrix*. It is a square matrix containing logical values (true or false). The size of a matrix is equal to the number of nodes. Let's assume that nodes are labelled with consecutive integers starting from 0. If node *u* links (there exists a hyperlink) to the node *v*, the value of the matrix in a row *u* and column *v* is *true*. Otherwise it is *false*. One may observe that for an undirected graph the matrix is symmetric. An excerpt of a matrix is shown in figure 2.

---

http://pages.cpsc.ucalgary.ca/~jacobs/Courses/cpsc331/W12/tutorials/Figures/directed_graph_exampl
e1.gif [May, 2012]
7 http://www.worldwidewebsize.com/ [May 15, 2012]

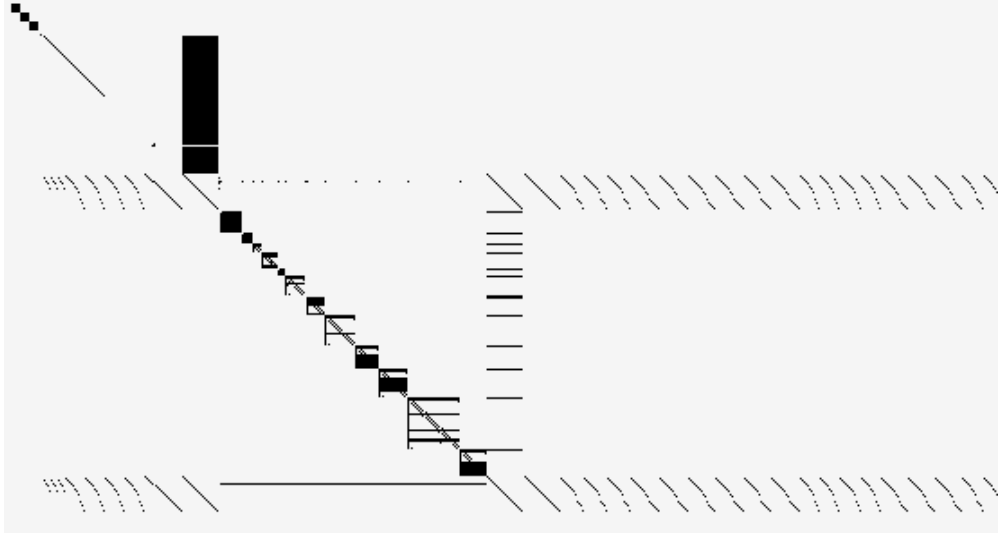

*Figure 2. An excerpt of an adjacency matrix*[8]

Such a representation is extremely fast at answering queries for adjacency between two arbitrary nodes (it requires constant time). However, in order to retrieve the list of all successors of a node one has to iterate through the whole row checking for adjacency. The same rule applies to retrieving the list of predecessors with the difference of checking the whole column instead of a row. In both cases we have to check as many fields as the number of nodes for each query.

Another feature of an adjacency matrix that is worth mentioning is its relatively high memory consumption, which is not a problem when the graph is dense (all known techniques storing graph in an uncompressed way have their memory cost similar to the adjacency matrix). However, if the matrix is sparse, there exist other methods consuming less resources. One of them is described in the following paragraph.

---

[8] Snapshot of EU-2005 taken from WebGraph Explorer – software written for Bachelor's thesis in [2]

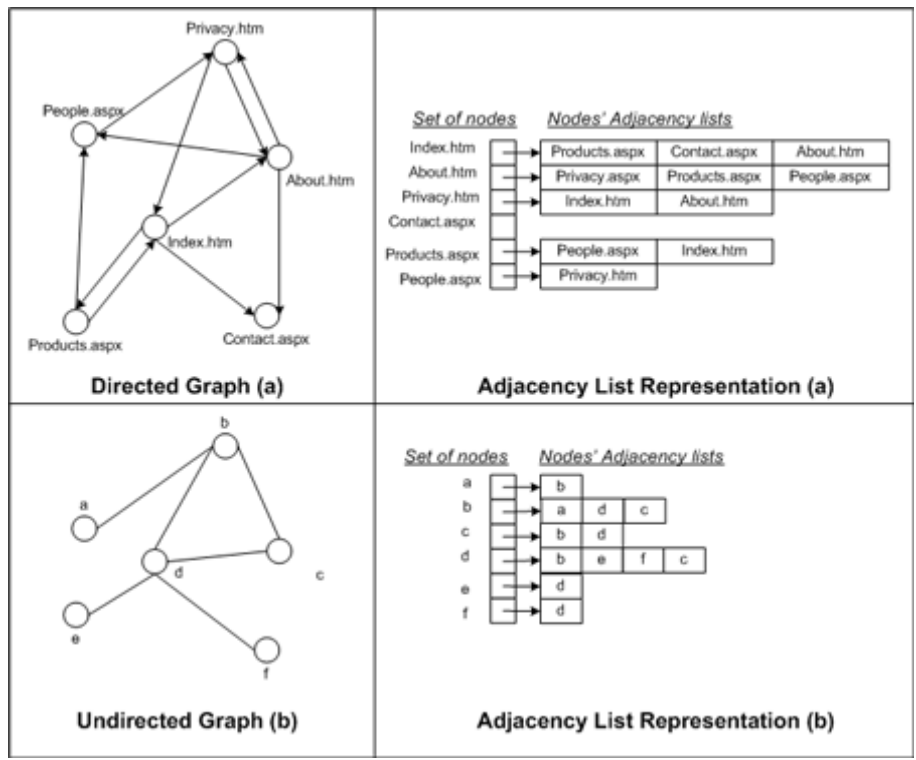

*Figure 3. An example of a simple web graph with its adjacency list[9]*

An *adjacency list* is an array of lists of successors of each node. An example is shown in figure 3. Each successor denotes an edge, so the list's memory consumption is proportional to the number of edges in a graph. For sparse graphs adjacency lists utilize much less memory than adjacency matrices. Moreover, if one needs to retrieve the list of successors of a specific node it can be done in linear time with respect to number of successors of that node. However, in order to obtain the list of predecessors we have to iterate through the whole graph to count all the edges pointing to the specific node. Checking adjacency of two vertices is also slower than using a matrix, because it involves looking at each successor of the node.

While comparing techniques of web graph compression, there are two main factors: the access time (which is usually a measure of time needed to obtain the list of successors of a specific node or needed to acquire one edge) and compression ratio (that is related to space needed to store the link). All ideas proposed in this work are compared to other methods using those quantities. There are other less important aspects as compression time and memory usage. They are quite significant since

---

the Internet changes constantly and the graphs are being updated all the time, however, it is not that essential as two main factors. Figure 4 shows a sample diagram used for methods comparison.

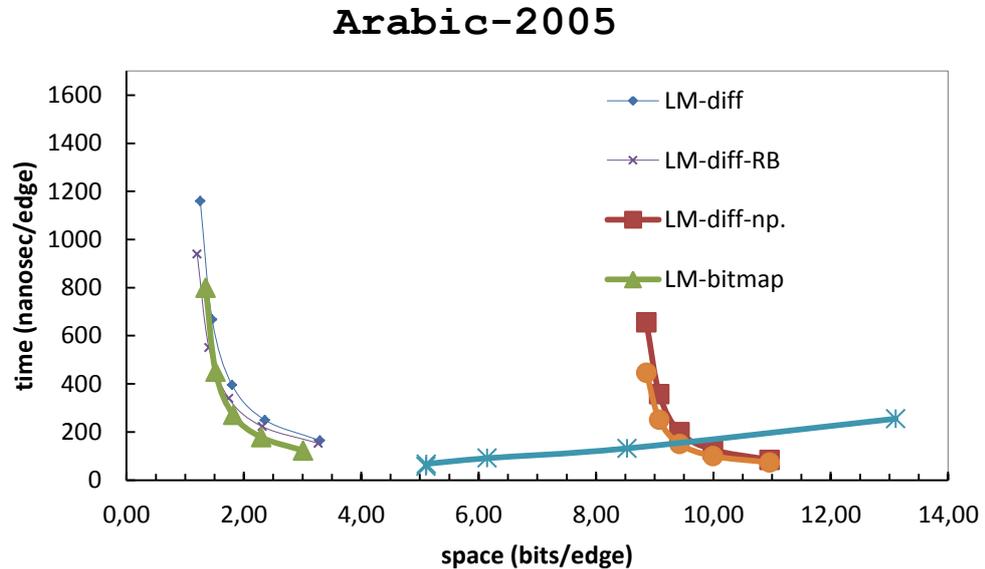

*Figure 4. An example of a diagram showing random access time with respect to compression ratio for different methods*[10]

While mentioning different compression techniques it is also worth noting that when we refer to 1D algorithms it means that such a method is efficient in obtaining successors lists only, while 2D algorithms are designed to provide fast access to both successors as well as predecessors of a node. Most of the time when both functionalities are necessary it is convenient to prepare the transpose of a graph and store it using 1D method, which is usually faster than current 2D techniques. As a result each successor in such transposed matrix specifies a predecessor in normal conditions.

---

[10] Picture taken from [12]

# 3. PROBLEM DESCRIPTION

## 3.1. The reason for compressed formats

As mentioned in the previous chapter the size of the Web was estimated to 45 billion pages. Let's assume that each page contains 25 links on average. This gives us a graph of 45 billion nodes and over 1 trillion edges. If we were to use one of the techniques above we would need around $2.5 \times 10^{20}$ bytes (250 EB) or $9 \times 10^{12}$ bytes (9 TB) respectively, which is out of reach for majority of contemporary machines. Although there exist distributed systems that would process such a huge representation, the network communication between them is a large bottleneck being slower even than a hard disk in comparison [1]. This exposes the need for much more compact graph representation that would fit into the memory and provide competitive access time. There are many projects emerging nowadays trying to achieve the best time in obtaining random node's successor list while maintaining high compression ratio. Boldi and Vigna have a great summary of currently available techniques in [3]. Let us present some of them.

## 3.2. *WebGraph* framework

The authors of the same article have been working on web graphs for years and have collected great deal of information about them, thus they provide one of the biggest source of information about the nature of those graphs. A web crawler developed at *Laboratory for Web Algorithms* (*LAW*)[11], called *UbiCrawler*, helped much in collecting data for the experiments which now serve as a major testing material in experiments on new web graph compression techniques (they are going to be used in this thesis as well). More information about the sets will be presented in chapter 6.

Starting from the early idea drafts back in 2003 [1] they developed the whole framework for creating, storing and exploring web graphs written in Java, which is the most popular technology used by the authors of techniques presented in this thesis. Their idea is based on a simple observation, that most of the times the adjacency list of the node nearby differs only in small aspect from the list of the current one (assuming

---

[11] http://law.dsi.unimi.it/ [May, 2012], Boldi and Vigna constitute half of its members

nodes are ordered in a lexicographical manner with respect to corresponding URLs). This is due to the fact that many subpages of one domain have common structure and point to similar pages. Exploiting this fact one can represent the list of successors of a node as a list of differences between current adjacency list and one from the lists processed earlier. This can be done by storing the id of the node whose list is being copied, an array of 0's and 1's stating whether successor is found in the current list and additional nodes that were not present in the previous list. Their method goes even further by compressing those copy lists of 0's and 1's. It is achieved by saving number of consecutive 0's decreased by one and then number of consecutive 1's decreased by one, etc. In this manner we obtain a representation that for each node keeps an information about the reference adjacency list, number of copy blocks, copy blocks themselves and extra successors. Now those successors may be presented as intervals of differences between successive nodes, which is the case. Specifically each interval is stored as a smallest value in the interval and its length decreased by 2 (interval contains at least two values) and each of the remaining nodes is compressed using differences.

| Node | Outd. | Ref. | # blocks | Copy blocks | Extra nodes |
|------|-------|------|----------|-------------|-------------|
| . . . | . . . | . . . | . . . | . . . | . . . |
| 15 | 11 | 0 | | | 13, 15, 16, 17, 18, 19, 23, 24, 203, 315, 1034 |
| 16 | 10 | 1 | 7 | 0, 0, 2, 1, 1, 0, 0 | 22, 316, 317, 3041 |
| 17 | 0 | | | | |
| 18 | 5 | 3 | 1 | 4 | 50 |
| . . . | . . . | . . . | . . . | . . . | . . . |

*Figure 5. Final BV scheme adjacency list structure*[12]

Summing up, one list of successors of each node is represented by the outdegree of a node, reference number (the difference between current node number and the number of a node, whose list is being copied), count of copy blocks, copy blocks themselves, number of intervals, smallest values of each interval (its left extremes), interval lengths and finally compressed list of residuals. This is shown in figure 5. One can notice that the length of the list of residuals is not known, but can be easily obtained from the difference between outdegree and number of successors enclosed in copy blocks. Such complicated structure may be questionable in terms of compression rate, but results in very tolerable outcome in practice. Very commonly used (in web graph studies) crawl

---

[12] Picture taken from [1]

of .uk domain from 2002 called *UK-2002* containing 18 520 486 nodes and 298 113 762 edges can be put into 107 MB file[13] giving 3.01 bits/link compression, which is totally acceptable. It is stated that the same file can be compressed even to 1.805 bits/link, but such a representation can be accessed sequentially only. Another technique that was developed in LAW called *Layered Label Propagation* (*LLP* described in [4]) allowed to reorder nodes in such a way that random access is still possible and the compression ratio improved to 2.28 bits/link. Still this method needs to store the additional information about nodes (their URLs), so we cannot compare this method to others since the input and output of them differs significantly. Furthermore, the reordered list is more complex to be sought through and there are no tests performed claiming the compression ratio and access time after the reorder. However, it is worth to notice that reordering nodes can make the whole graph representation even smaller.

The main parameters of the so-called BV graph representation are window size $W$ (the maximum number of nodes we look back in order to choose one for its adjacency list to be copied), minimum interval length $L_{min}$ (sometimes it is better to store differences between residuals than creating little intervals) and the maximum reference count $R$ (since adjacency lists are copied from previous ones, the reference chain may occur and this value controls maximum number of jumps). The first parameter obviously controls compression time, since there more successor lists to analyse while choosing the one to be copied for a node. Minimum interval length gives best compression ratio if set to 3 [2]. The last parameter gives better compression ratio when increased at a cost of access time, because with higher amount of references more lists are likely to be iterated while retrieving list of successors and that consumes time. It can be seen in table 1 taken from [1].

---

| 18.5 Mpages, 300 Mlinks from .uk | | | | | | | | |
|---|---|---|---|---|---|---|---|---|
| R | Average reference chain | | | Bits/node | | | Bits/link | | |
| | $W=1$ | $W=3$ | $W=7$ | $W=1$ | $W=3$ | $W=7$ | $W=1$ | $W=3$ | $W=7$ |
| $\infty$ | 171.45 | 198.68 | 195.98 | 44.22 | 38.28 | 35.81 | 2.75 | 2.38 | 2.22 |
| 3 | 1.04 | 1.41 | 1.70 | 62.31 | 52.37 | 48.30 | 3.87 | 3.25 | 3.00 |
| 1 | 0.36 | 0.55 | 0.64 | 81.24 | 62.96 | 55.69 | 5.05 | 3.91 | 3.46 |
| Tranpose | | | | | | | | | |
| $\infty$ | 18.50 | 25.34 | 26.61 | 36.23 | 33.48 | 31.88 | 2.25 | 2.08 | 1.98 |
| 3 | 0.69 | 1.01 | 1.23 | 37.68 | 35.09 | 33.81 | 2.34 | 2.18 | 2.10 |
| 1 | 0.27 | 0.43 | 0.51 | 39.83 | 36.97 | 35.69 | 2.47 | 2.30 | 2.22 |
| 118 Mpages, 1 Glinks from WebBase | | | | | | | | | |
| R | Average reference chain | | | Bits/node | | | Bits/link | | |
| | $W=1$ | $W=3$ | $W=7$ | $W=1$ | $W=3$ | $W=7$ | $W=1$ | $W=3$ | $W=7$ |
| $\infty$ | 85.27 | 118.56 | 119.65 | 30.99 | 27.79 | 26.57 | 3.59 | 3.22 | 3.08 |
| 3 | 0.79 | 1.10 | 1.32 | 38.46 | 33.86 | 32.29 | 4.46 | 3.92 | 3.74 |
| 1 | 0.28 | 0.43 | 0.51 | 46.63 | 38.80 | 36.02 | 5.40 | 4.49 | 4.17 |
| Tranpose | | | | | | | | | |
| $\infty$ | 27.49 | 30.69 | 31.60 | 27.86 | 25.97 | 24.96 | 3.23 | 3.01 | 2.89 |
| 3 | 0.76 | 1.09 | 1.31 | 29.20 | 27.40 | 26.75 | 3.38 | 3.17 | 3.10 |
| 1 | 0.29 | 0.46 | 0.54 | 31.09 | 29.00 | 28.35 | 3.60 | 3.36 | 3.28 |

*Table 1. Experimental data using BV scheme about reference chains with $L_{min} = 3$.*

Based on this idea the whole framework has been written and published on the website of LAW[14]. The code is available and it includes the graph class itself, iterators, adapters to be able to use it for specific purposes and some examples. The manual is added to the packet. It is automatically generated from JavaDoc[15] and contains description of all the classes contained in the framework. Attempts have been made to port the project to another programming language like C++ or Python and their outcomes can be downloaded from the same web page.

Thanks to its satisfactory compression ratio and access time as well as the fact that this idea is being used for generating biggest crawls available online, it became a standard in web graph compression. That means that every new method has to be compared with this one in ordered to be publicly approved (at least in the sense of random 1D access).

---

[14] http://webgraph.dsi.unimi.it/ [May, 2012]
[15] http://www.oracle.com/technetwork/java/javase/documentation/index-jsp-135444.html [May, 2012]

### 3.3. Virtual Node Miner

In spite of having a standard in web graph compression, people still seek for other solutions. One of the examples is an idea of compression using community detection [5]. The main problem with adjacency lists is that their size depends on the number of edges. So in order to reduce this number the following method has been developed. If there exists bipartite strongly connected directed sub-graph (each one of $m$ vertices has $n$ edges directed to other $n$ vertices), we can introduce the new supernode that links to all nodes being linked to in the previous configuration and which all the previously directing nodes direct to. This is shown in figure 6 taken from [5]. In that manner we obtain a sub-graph with one additional node and number of edges reduced by $m \times n - (m + n)$. For only 10 nodes directing and 5 being directed to this gives reduction of (50-15) / 50 = 70%.

As tests show there is plenty of such bipartite cliques so that the overall compression ratio is very satisfactory and beats even the BV scheme. However, the work is not fully tested and the code is not available so it is not easy to determine whether this method depends on the order of the input. Due to this fact it was impossible to acknowledge this method and compare the new solution to that algorithm.

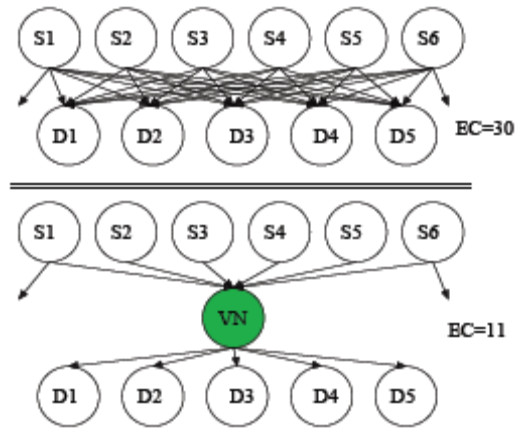

*Figure 6. Inserting virtual node in order to reduce number of edges*[16]

---

[16] http://www.wsdm2009.org/wsdm2008.org/WSDM2008-papers/p95.pdf [May, 2012]

### 3.4.Breadth First Search

Another approach can be used in order to compress a web graph. In [6] there was presented an innovative and very interesting method of traversing graph using Breadth First Search (BFS) in order to obtain successive nodes whose lists are to be compressed thereafter. The idea is very simple: one has to traverse through the whole graph labelling each node with the number of its successors that have not been visited and are not linked by visited nodes. Then the queue is constructed using so-called *traversal lists* and the lists of successive nodes are coded using one out of four types of coding. What is innovative here is the fact, that two types depend on the same successors counting from left of the previous list. That type of similarity has not been exploited in none of the methods presented in this thesis. The BFS search also helps exploiting locality of the edges (usually there are much more links within the same domain, called *intralinks*, than links pointing to other nodes - *interlinks*). However, the outcome is based on the node chosen for the beginning of the traversal, thus the results can differ much when other way of choosing nodes is being implemented or tests are slightly modified.

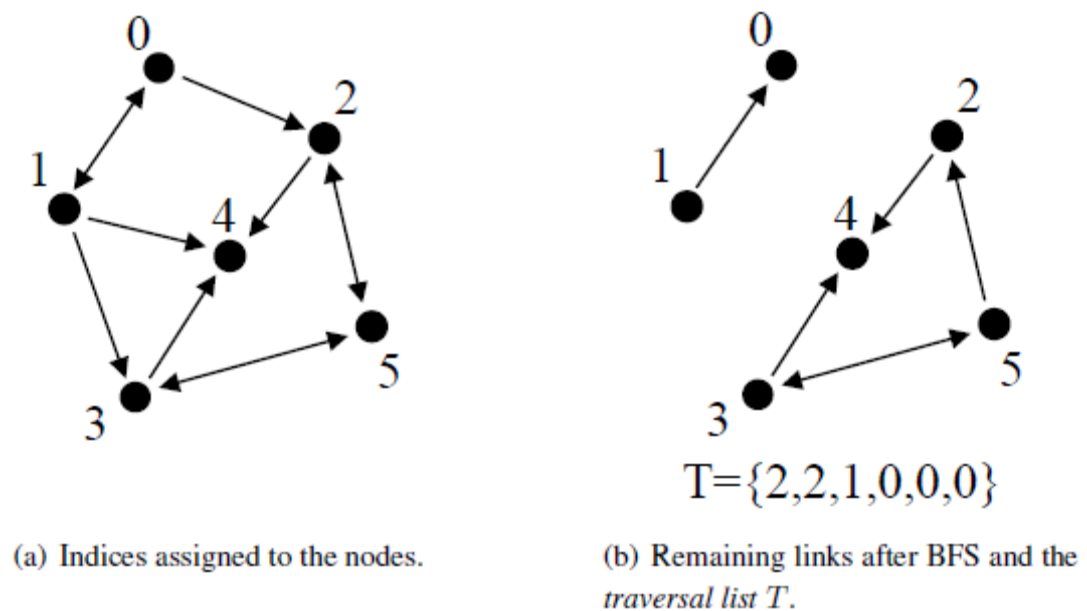

(a) Indices assigned to the nodes.

(b) Remaining links after BFS and the traversal list $T$.

*Figure 7. Illustration of traversing through web graph using BFS[17]*

---

[17] Picture taken from [6] available at http://www.mdpi.com/1999-4893/2/3/1031/pdf [May, 2012]

The algorithm is divided into two phases. Phase 1, presented in figure 7, involves traversing through the whole graph while the second phase is responsible for compressing adjacency lists. There were some observations made in order to improve compression. First, using this technique two linked nodes are likely to have close identifier. What is more, the consecutive adjacency lists share many successors since typically two linked documents contain mostly the same links (it is worth noticing that this differs from the similarity of nodes sorted using natural order, because here we have similarity among linked documents, thus not necessarily in the same domain).

Exploiting those facts the method makes use of four types of redundancies:

1. There exist successive nodes having same adjacency lists, so there is only one list with number expressing the amount of nodes sharing same list.

2. Usually the difference between outdegrees of consecutive nodes is very small, so they are gap-encoded.

3. Sometimes there are consecutive successors, so they are encoded as an interval if their length is smaller than $L_{min}$ parameter.

4. If the consecutive lists have identical consecutive successors and such a box is bigger than $A_{min}$ parameter, then they are run-length encoded.

| Degree | Links |
|---|---|
| ...<br>0 | ... |
| 9 | $\beta\,7$   $\phi\,1$   $\phi\,1$   $\phi\,1$   $\phi\,0$   $\phi\,1$   $\phi\,1$   $\phi\,1$   $\phi\,1$ |
| 9 | $\beta\,0$   $\beta\,1$   $\beta\,0$   $\beta\,0$   $\beta\,0$   $\beta\,0$   $\beta\,0$   $\beta\,0$   $\beta\,2$ |
| 10 | $\beta\,0$   $\beta\,1$   $\beta\,0$   $\beta\,0$   $\beta\,0$   $\beta\,0$   $\beta\,0$   $\beta\,0$   $\beta\,1$   $\phi\,903$ |
| 10 | $\beta\,0$   $\beta\,1$   $\beta\,0$   $\beta\,0$   $\beta\,0$   $\beta\,0$   $\beta\,0$   $\beta\,0$   $\beta\,223$   $\phi\,900$ |
| 10 | $\beta\,0$   $\beta\,1$   $\beta\,0$   $\beta\,0$   $\beta\,0$   $\beta\,0$   $\beta\,0$   $\beta\,0$   $\beta\,1$   $\alpha\,0$ |
| 10 | $\beta\,0$   $\beta\,1$   $\beta\,0$   $\beta\,0$   $\beta\,0$   $\beta\,0$   $\beta\,0$   $\beta\,0$   $\beta\,1$   $\beta\,0$ |
| 10 | $\beta\,0$   $\beta\,1$   $\beta\,0$   $\beta\,0$   $\beta\,0$   $\beta\,0$   $\beta\,0$   $\beta\,0$   $\beta\,1$   $\beta\,0$ |
| 10 | $\beta\,0$   $\beta\,1$   $\beta\,0$   $\beta\,0$   $\beta\,0$   $\beta\,0$   $\beta\,0$   $\beta\,0$   $\beta\,1$   $\beta\,0$ |
| 10 | $\beta\,0$   $\beta\,1$   $\beta\,0$   $\beta\,0$   $\beta\,0$   $\beta\,0$   $\beta\,0$   $\beta\,0$   $\beta\,1$   $\beta\,0$ |
| 10 | $\beta\,0$   $\beta\,1$   $\beta\,0$   $\beta\,0$   $\beta\,0$   $\beta\,0$   $\beta\,0$   $\beta\,0$   $\alpha\,76$   $\alpha\,232$ |
| 9 | $\beta\,0$   $\beta\,1$   $\beta\,0$   $\beta\,0$   $\beta\,0$   $\beta\,0$   $\beta\,0$   $\beta\,0$   $\beta\,0$ |
| ... | ... |

*Figure 8. Sample adjacency lists compressed with redundancy reduction.*
*Each encoding consists of the integer gap between adjacent elements in the list and a type indicator chosen in the set {α,β,χ,φ} needed in decoding.*[18]

There is an example shown in figure 8 presenting example lists being compressed using previously mentioned redundancy reduction. Although being quite complex, the method provides excellent compression ratios, which are often the best among all known algorithms at the same time having fast access to nodes. However, the nodes have to be reordered in the representation, which makes it hard to map resulting successors with URLs (it is necessary to store the list of URLs together with ids of nodes, which is not space consuming when URLs are naturally ordered which is not the case). In the next section another method that can make use of the similarities in the consecutive rows is described.

### 3.5. Compression of patterns

In [7] Asano et al. defined six types of blocks on adjacency matrix. They use it for intra-host links (nodes are naturally ordered as usual), while inter-host ones are

---

renumbered in order to obtain another adjacency matrix which is then compressed using previously mentioned six techniques. They consist of:

- one isolated element is coded by a singleton block,

- horizontal line of two or more edges create a horizontal block,

- vertical line of two or more edges create a vertical block,

- if a horizontal or vertical line construct an L-shaped figure it is stored with the upper left element,

- if there two or more consecutive horizontal lines of the same length covering the same edges it is saved as a rectangular block,

- there is a diagonal block of consecutive elements from consecutive lists.

There is an illustration in [7] explaining those 6 techniques using the example matrix. It is shown in figure 9.

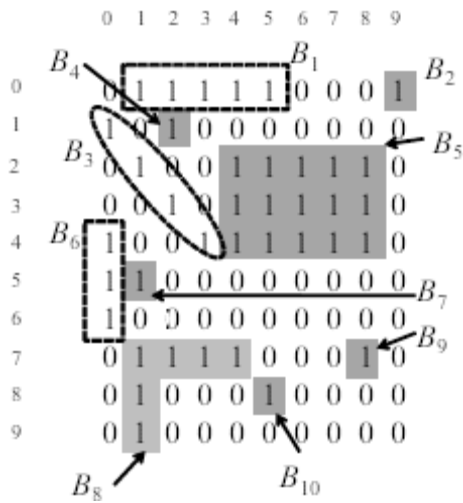

*Figure 9.Illustration found in **[7]** demonstrating six redundancies found in an adjacency matrix.*

Such a representation provides the better compression ratio than BV scheme (at least for the sets tested, that are quite small), but as one can imagine the time needed to obtain the whole list of successors is several orders of magnitude apart from the other solutions as they can depend on so many blocks, that it is very time consuming to extract all of them. Thus despite excellent compression ratio this method cannot be classified as a good web graph compressing technique. Still it is worth mentioning.

### 3.6. Re-Pair edge compression

Although we can change the order of nodes like in BFS example, one may think about compressing the links themselves. In [9, 10], the Re-Pair idea is being used to compress adjacency lists by searching for the most frequently used pair of successors, replacing them by the new symbol and adding that modification to a dictionary until it becomes less convenient to do so. Sample example is provided in figure 10.

| Pair | String |
|------|--------|
| | `singing.do.wah.diddy.diddy.dum.diddy.do` |
| A → .d | `singingAo.wahAiddyAiddyAumAiddyAo` |
| B → dd | `singingAo.wahAiByAiByAumAiByAo` |
| C → Ai | `singingAo.wahCByCByAumCByAo` |
| D → By | `singingAo.wahCDCDAumCDAo` |
| E → CD | `singingAo.wahEEAumEAo` |
| F → in | `sFgFgAo.wahEEAumEAo` |
| G → Ao | `sFgFgG.wahEEAumEG` |
| H → Fg | `sHHG.wahEEAumEG` |

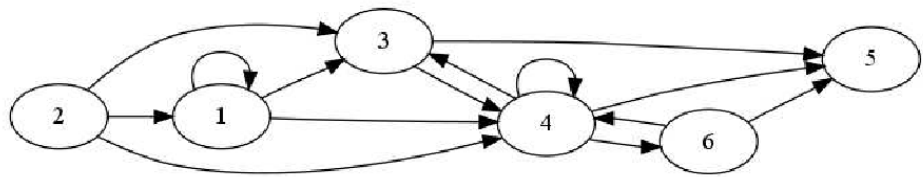

*Figure 10. Recursive pairing in practice.*[19]

To use this method in web graphs all lists are concatenated with unique negative identifiers in order to be able to distinguish different lists during decompression. The idea seems to be simple yet gives quite satisfactory results. For EU-2005 it yields 4.68 bits/link. It is worth to mention that method is suited for web graphs, since it permits fast local decompression, which is true in case of web graphs, handles well large alphabets, symbols that are unique will not be replaced, that is especially important while delimiting adjacency lists. What is more, Re-Pair can take advantage of partial similarity of two consecutive lists even if they are not identical.

---

[19] The picture is taken from the article [14]

Although being very promising the method is impractical to be used. The main reason for that is the huge memory consumption for the compression stage since we need to represent all edges in the memory in order to find same pairs. In case of bigger sets it is impossible, so as a result such sets are divided into smaller parts and then sought for most frequent pairs. One can notice that the best variants for one part may be not the optimal for the whole graph. For that reason this method is quite impractical, which is shown by the test results in [3].

### 3.7. $k^2$ partitioned

The new idea was presented using Re-Pair and other structure called $k^2$-*tree*, that is described in [11]. The $k^2$-*trees* are structures storing information about underlying surface. In details they are trees that represent squares that divide adjacency matrix. Each square is divided into $k^2$ smaller ones if there is at least one edge within the adjacency matrix. If not it is left as it is. In the tree this dependence is stored as 1 or 0 respectively and since 1 squares are divided, all leaves with 1's have another set of leaves to indicate the emptiness of each sub-square. There is one biggest square of the size divisible by $k^2$ which is represented by the root of the tree. The division of a matrix is shown in figure 11.

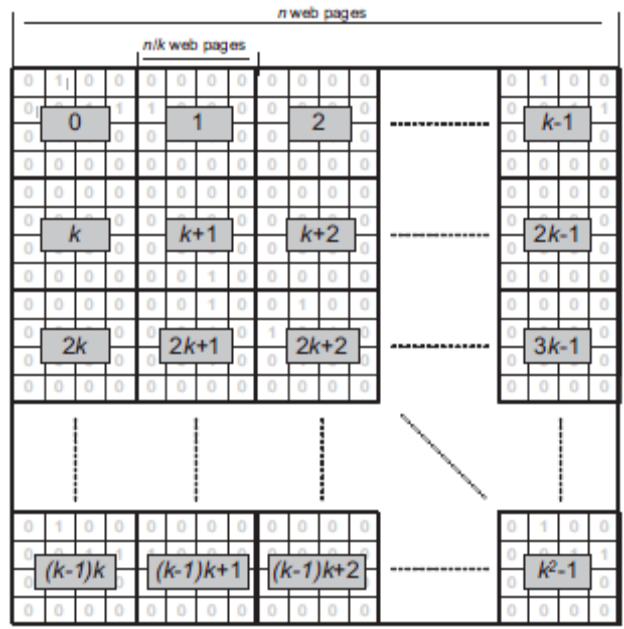

*Figure 11.Adjacency matrix being divided into $k^2$ trees.*[20]

It would be possible to represent the whole graph in this manner, as it is presented in figure 12, and there were tests conducted in [2] showing quite satisfactory results especially regarding compression ratio, far from the best though. That is why the next step has been made and in spite of representing the whole structure as a tree, the smaller squares that are most likely to be representing some neighbouring nodes in terms of locality (inner links) are compressed using *Re-Pair* technique. That allows achieving much higher ratios at the same time maintaining satisfactory access time, still longer than BV scheme. The method requires URLs to be sorted in a lexicographical manner, since its main compressing factor comes from the locality of the nodes.

---

[20] The picture is taken from the article [11]

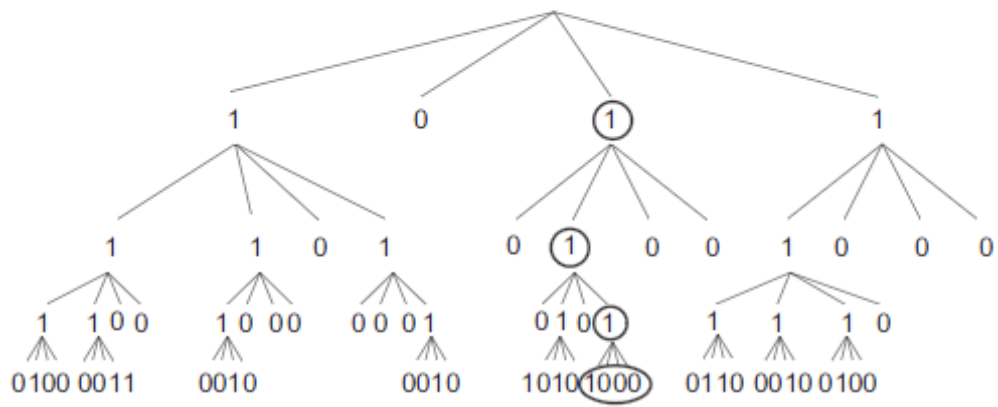

| 0 | 1 | 0 | 0 | 0 | 0 | 0 | 0 | 0 | 0 | 0 | 0 | 0 | 0 | 0 | 0 |
|---|---|---|---|---|---|---|---|---|---|---|---|---|---|---|---|
| 0 | 0 | 1 | 1 | 1 | 0 | 0 | 0 | 0 | 0 | 0 | 0 | 0 | 0 | 0 | 0 |
| 0 | 0 | 0 | 0 | 0 | 0 | 0 | 0 | 0 | 0 | 0 | 0 | 0 | 0 | 0 | 0 |
| 0 | 0 | 0 | 0 | 0 | 0 | 0 | 0 | 0 | 0 | 0 | 0 | 0 | 0 | 0 | 0 |
| 0 | 0 | 0 | 0 | 0 | 0 | 0 | 0 | 0 | 0 | 0 | 0 | 0 | 0 | 0 | 0 |
| 0 | 0 | 0 | 0 | 0 | 0 | 0 | 0 | 0 | 0 | 0 | 0 | 0 | 0 | 0 | 0 |
| 0 | 0 | 0 | 0 | 0 | 0 | 0 | 0 | 0 | 0 | 0 | 0 | 0 | 0 | 0 | 0 |
| 0 | 0 | 0 | 0 | 0 | 0 | 1 | 0 | 0 | 0 | 0 | 0 | 0 | 0 | 0 | 0 |
| 0 | 0 | 0 | 0 | 0 | 0 | 1 | 0 | 0 | 1 | 0 | 0 | 0 | 0 | 0 | 0 |
| 0 | 0 | 0 | 0 | 0 | 0 | 1 | 0 | 1 | 0 | 1 | 0 | 0 | 0 | 0 | 0 |
| 0 | 0 | 0 | 0 | 0 | 0 | 1 | 0 | 0 | 1 | 0 | 0 | 0 | 0 | 0 | 0 |
| 0 | 0 | 0 | 0 | 0 | 0 | 0 | 0 | 0 | 0 | 0 | 0 | 0 | 0 | 0 | 0 |
| 0 | 0 | 0 | 0 | 0 | 0 | 0 | 0 | 0 | 0 | 0 | 0 | 0 | 0 | 0 | 0 |
| 0 | 0 | 0 | 0 | 0 | 0 | 0 | 0 | 0 | 0 | 0 | 0 | 0 | 0 | 0 | 0 |
| 0 | 0 | 0 | 0 | 0 | 0 | 0 | 0 | 0 | 0 | 0 | 0 | 0 | 0 | 0 | 0 |
| 0 | 0 | 0 | 0 | 0 | 0 | 0 | 0 | 0 | 0 | 0 | 0 | 0 | 0 | 0 | 0 |

*Figure 12.Accessing the adjacency matrix using $k^2$ trees.*[21]

Despite the fact of being slower this method allows to store both successors as predecessors of nodes at the same time (previous methods required storing transposed graph in order to be able to output the list of predecessors, which doubled the compression ratio). So this actually is the first 2D technique of web graph compression and as the idea presented in this thesis is also a 2D algorithm this is the best opponent to be chosen for comparison.

---

[21] The picture is taken from the article [11]

### 3.8. List Merging

Yet another method exploiting similarity of local nodes was proposed in [10]. The idea is quite similar to BV scheme although some changes occur. First of all the whole graph is divided into chunks of nodes of size equal to some power of two. Each chunk is then compressed using following methodology: first, the list of all unique successors appearing in the lists of chunk is created. Then for each such unique successor we check whether it exists in each list putting flag 1 if it is true and zero otherwise (for number of rows in the chunk divisible by eight this information will fill whole byte; this parameter should be additionally the power of two in order to enable the use of bitwise operations, which are faster than arithmetical ones). Last, the list of unique successors is differentially encoded and then byte coded, while the array of flags is compressed using Deflate algorithm. The whole process is presented in figure 13. The graph is represented by a buffer filled with values created in a manner presented and the array of offsets to the positions in the buffer (one can notice, that the offsets are necessary for random access only).

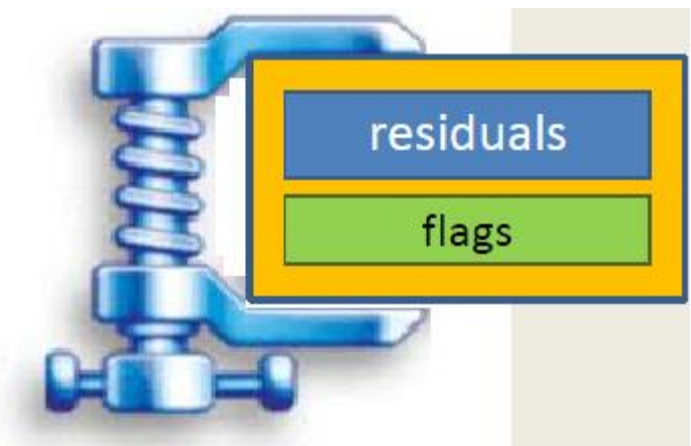

*Figure 13. List Merging in practice.*[22]

This 1D algorithm proved to have better compression ratio than BV scheme, but with slower access as a drawback. This is due to the characteristics of Deflate algorithm which is very efficient at compressing data, but relatively slow. Surprisingly such simple algorithm shows better characteristics regarding compression ratio multiplied by access time (it is faster while maintaining same size or smaller with same access time) compared to other methods presented earlier.

---

[22] The picture is taken from the materials used in presentation during conference [11]

There existed another version previously (called LM-diff) described in [13] storing flags as gaps between consecutive 1's. However, tests showed that this method gives worse results in access time and insufficient compression increase. Thus when referring to LM in this paper the LM-bitmap is being chosen.

# 4. SOLUTION

This chapter attempts to expose successive steps in the project timeline from the beginning debate until final test procedure.

## 4.1. First step

As stated in the introduction when the execution time and efficient resource management count, the C++ programming language is unbeatable among most popular languages, i.e. Java or Python. Thus if we wanted to obtain faster method we would need to have some base written in C++ to compare to. Moreover the tests ought to be conducted on the same machine for the comparison to be reliable. Most of the algorithms presented in the previous chapter were coded in Java or their code was unavailable. Hence the idea of rewriting the most promising ones into C++ arose so as to be able to compete with the best methods available on the same level.

Starting with the method with the lowest compression ratio and access times the process to code LM idea in C++ began. Its main principle is shown in code snippet 1.

```
1    outF ← [ ]
2    i ← 1
3    for line_i, line_{i+1}, …, line_{i+h-1} ∈ G do
4        tempLine_1 ← line_i ∪ line_{i+1} ∪ … ∪ line_{i+h-1}
5        tempLine_2 ← removeDuplicates(tempLine_1)
6        longLine ← sort(tempLine_2)
7        items ← diffEncode(longLine) + [0]
8        outB ← byteEncode(items)
9        for j ← 1 to |longLine| do
10           f[1…|longLine|] ← [0,0,…,0]
11           for k ← 1 to h do
12               if longLine[j] ∈ line_{i+k-1} then f[k] ← 1
13           append(outF, bitPack(f))
14       compress(concat(outB, outF))
15       outF ← [ ]
16       i ← i + h
```

*Code snippet 1. Pseudo-code of GraphCompressLM(G,h).* [23]

Since the method utilizes Deflate algorithm it was necessary to include $zlib$ [24] library. It was configured not to include header in order to decrease the size of the output. Also the best compression flag has been set. The decision was made to include methods for loading web graph from a text file, from previously computed files, storing graph to files or to a text file and return the list of successors of a given node.

---

[23] Code taken from [13]

The web graph uncompressed representation was stored in text files with number of lines equal to number of nodes and consecutive successors of a node separated by space character within the line. The nodes were assigned integer values from 0 to the total number of nodes decreased by one. Thus successor numbers were also integers specifying the node being linked to.

### 4.2. LM compression

The compressor for C++ performs almost same actions as its Java equivalent. However, there are some minor differences in the way temporal data is stored. Both options read number of lines first and then starts to process the file. Java version reads input line by line and processes it instantly. The second version uses input buffer since reading the whole chunk of a file and then splitting it in the RAM is faster than reading text lines. It is also impossible to read the whole file to memory at once since graph files are usually bigger than the memory available. Even when the file could be put into memory, it used that much space, that swapping occurred (operating system had to put a part of memory of other threads into the disk) resulting in worse performance than while using read buffer.

Now the lines are processed one by one filling the array of residues and temporal lists (specifically copying them). In our implementation the sorted array was used for residues (sorted using quicksort, whose complexity is $O(n\ log\ n)$ on average) and bit arrays for successor lists. It is hard to estimate sufficient length of the sorted array, which can contain duplicates if both successors appear in two different lines, because if in a chunk every node had a link to all others, it would take memory space proportional to number of nodes decreased by one and multiplied by chunk size. For Arabic-2005 set and chunk size equal to 128, it would take (22 744 080-1)*128*4 B = 11 644 968 448 B, which is over the capacity of nowadays PCs. Luckily tests show, that the length can be just a fraction of the biggest size possible. For example the same set has the longest list of successors equal to 9905 and transposed set has the longest list of successors equal to 575 618. Compared to the number of nodes in total (22 744 080) we obtain the ration of (575 618 * 128) / 22 744 080 = 3.23 assuming that all of lines in a chunk would be of a length of the longest one (which is highly improbable, due to the fact, that there are little or no such sites that are so close to each other in a lexicographical order and have so many pages that link to them). That's why it is sufficient to have an array of size equal to doubled number of nodes. The line processing operation is listed in code snippet 2.

---

[24] http://www.zlib.net/ [May, 2012]

```
while(readBuffer[readBufferStart] != '\n') {
    if(readBuffer[readBufferStart] == ' ') {
        sets[count][tmp>>3] |= 1 << (tmp & 7);
        allInts[allIntsSize++] = tmp;
        tmp = 0;
    } else tmp = (tmp<<3) + (tmp<<1) + readBuffer[readBufferStart] - '0';
    readBufferStart++;
}
readBufferStart++;
```

*Code snippet 2. Line processing loop.*

Bit arrays are of length equal to number of nodes divided by eight, since they have to contain all the references to all successors possible (one may observe, that they copy the whole row from adjacency matrix). Their role is to be able to answer the question whether this node is linked to arbitrary one as fast as possible. When number of lines processed is equal to number of lines specified for the chunk, the compression is being conducted. First, the array of residues is sorted and duplicates are being removed. Second, for all residues flags are being set for each node. Third, residues are being differentially and byte encoded and last, both residues and flags are being zipped using Deflate algorithm. The output is concatenated and after each concatenation the offsets are written to the offset table in order to provide random access.

Originally in Java residues are stored in the tree structure ($O(n \log n)$ in worst case, because the trees are balanced, where n equals to number of values inserted) rather than sorted array and the flags are being inserted into the array of hash sets. The same idea has been tried in C++ using STL yielding an unsatisfactory result. It occurred that equivalent structures (set and unordered_set respectively) are several times slower than in Java, which may be due to the STL overhead since it is general purpose and written with an aim of being universal.

Because of the fact that different structures have been used the compression time improved compared to the existing idea, however, it is not a subject of matter since it was not unsatisfactory earlier and what is more it is not the main factor of the method. More results are presented in chapter 6.

### 4.3. LM random access

Now trying to access the list of successors of a node we have to first calculate the correct chunk number, then locate its offset in the compressed graph structure and last unzip the whole chunk and decode its successors. One may observe that it involves iterating through the whole list of residues since it is differentially encoded and watching for the flag. If it is set to 1, then it means that current residue is a successor, so it is put into the result array.

The whole process is very fast (as experiments without using Deflate showed), however, zipping and unzipping takes fair amount of time, so the overall time is worse than in BV scheme. However, while creating C++ implementation few observations were made:

- Buffered input is much more efficient than loading whole file into the memory.

- Bit flags (or bitmaps), differential encoding and offsets are terribly fast structures.

- Deflate is very efficient in compression at a cost of being a bit slow for our purposes.

- When dividing the input into chunks it is possible to process each chunk separately, so there is space for multithreading support (for compression stage).

It is noticeable that although providing satisfactory compression rate and fast access this method is 1D only, so the plans were extended to cover 2D idea as well.

### 4.4.2D compression

The authors of LM technique developed such method that was based on the idea similar to $k^2$-*trees*, but instead of dividing squares into smaller ones only one level of recursion was used (this resulted in adjacency matrix being split into tiles of the same size noted as *B*). Next all the non-empty tiles, that can be seen in figure 14, are ordered and compressed using one of the four techniques:

- Each successor is assigned with the number of appearance if the tile was traversed horizontally from left to right row by row and the difference between consecutive successors is stored.

- The same method as in the previous point but zipped.

- Each successor is assigned with the number of appearance if the tile was traversed vertically from top to bottom column by column and the difference between consecutive successors is stored.

- The same method as in the previous point but zipped.

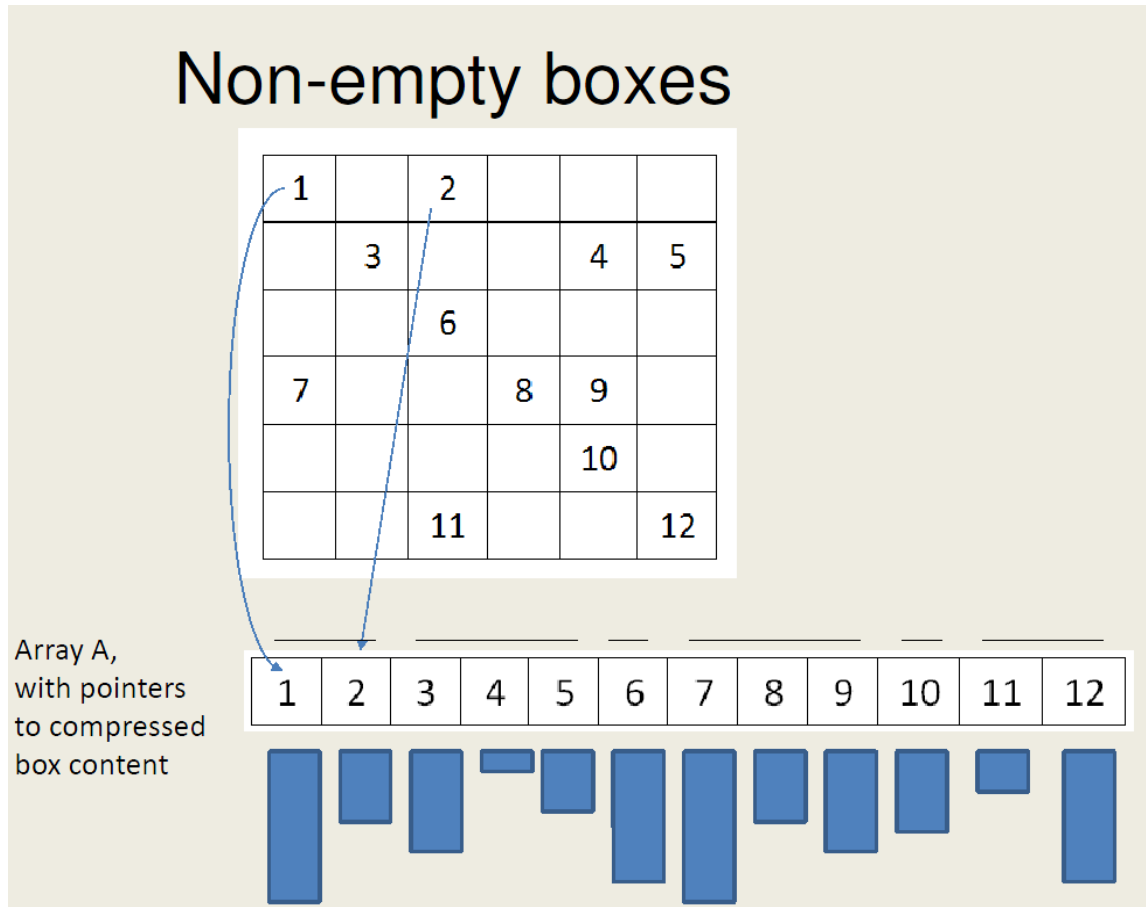

*Figure 14.2D compression in practice.*[25]

Now each non-empty tile while being compressed is treated with all of those methods and then the one that produces the smallest output is chosen (if two have the same smallest output, the unzipped version comes first and then the horizontal one). Last, the number of preceding tiles in a row and the number of preceding tiles in a column are added to the tile structure. The output is concatenated with the output of other tiles and the offsets are stored in two arrays. One array stores the offsets in row major order while the second stores indices to the first array in order to be able to determine the size of compressed tile data (tiles are being processed in a row major order, so the difference between horizontal offsets is the length of the compressed tile representation). Additionally the offset uses two most significant bits to encode compression type, that is which variant of the previously presented four methods has been chosen.

---

[25] The picture is taken from the materials used in the presentation during conference [11]

Such a structure allows to iterate through the graph sequentially, but since the random access is needed, it is necessary to delimit rows of tiles (and columns of tiles for the transposed matrix as well). Thus there is another array of offsets that are offsets of the first tiles in each row and second array of indices of the vertical offset array that correspond to first tiles in each column (again we cannot store offsets themselves because it would be impossible to determine the length of a compressed tile representation).

Altogether the structure holds compressed concatenated tiles representation, two arrays of offsets of all tiles (so their length is equal to number of nodes divided by tile size $B$ and squared) and two arrays of first offsets (so their length is equal to number of nodes divided by tile size $B$).

### 4.5.2D random access

In theory the idea introduces some trivial solutions (for example arrays of first offsets), but as practice showed the method produces satisfactory output of compression ratio being as low as 1.72 bits/link for EU-2005 dataset and $B = 1024$ or 0.98 bits/link for Indochina-2004 dataset and same $B$, which is one of the best compression ratios among all methods described in the previous chapter. However, the speed is totally unacceptable being two orders of magnitude slower than BV scheme for the same sets.

### 4.6.2D improvement

Despite quite slow access the compression ratio itself is quite impressive what resulted in pursue of a method to improve speed. One of the ideas that emerged was to introduce stripes that divide each tile stating whether there exist any successors (or predecessors for vertical stripes) for a given stripe. That would help to determine whether it is necessary to process the whole tile (which is costly) to find any successors or it can be omitted. Since the compression ratio is much better than in other methods we could afford adding additional data which were those stripes.

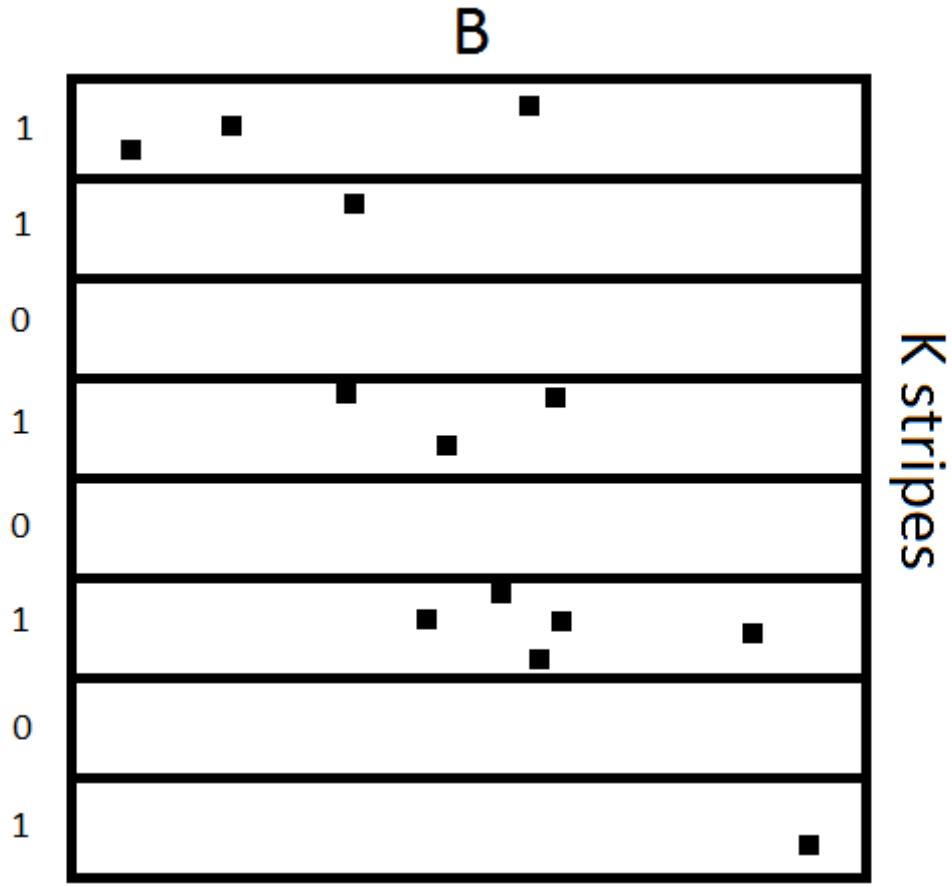

*Figure 15.2D tile divided by 8 stripes.*

Specifically there is a defined stripes amount, $K$, that determines how many stripes the tile is going to be divided into. The name "stripe" is used because when the square tile is divided into those regions of $B/K$ nodes it looks like it was striped, as seen in figure 15. Now each stripe is given 1 if there exists at least one edge in such region or 0 if the region is empty (there are no successors in this area). Due to the fact, that this is 2D algorithm we needed both horizontal stripes for successors and vertical stripes for predecessors.

The structure representing stripes is a bit array of length equal to stripes amount $K$. Because the bit array is byte-aligned both tile sizes $B$ and stripes amounts $K$ were chosen as powers of two. Since we need both kinds of stripes (i.e. horizontal and vertical), there are two such arrays increasing the size of each tile representation. The arrays are not being compressed or modified in any way, but just added to the tile structure.

# 5. IMPLEMENTATION

The application is written in C++ and contains three classes of objects:

- Class *CStripeGraph* representing graph, allowing its construction, loading from the text file or previously created StripeGraph files, saving them or writing text representation, successors and predecessors lists retrieval and showing some statistics.

- Class *CUncompressedTile* having methods for inserting nodes into the tile, compressing the tile itself using all four methods and then choosing the final one and storing information about stripes.

- Class *CCompressedTile* used for loading compressed tile representation, decoding data contained inside and successors or predecessors retrieval

All projects written for the purpose of this thesis were produced and compiled in Microsoft Visual Studio 2010 using Release x64 configuration. They both use *zlib* library in version 1.2.6 compiled using same application and configuration.

General purpose of the program was to compress graphs stored in text format, save them in a compressed structure to be used later for testing, test those representations using different parameters for compression ratio and access time and to write the results into some log files. It consists of seven files, six of which are headers and implementations of methods of three classes presented above. The whole testing procedure was coded into the main function of the program, which is located in the seventh file *stripes.cpp* which contains the entry point of the application. There were no command line arguments used in this project, so two executable files were created: one for compressing text representation and creating compressed files and second for testing. The paths to specific folders are also hard coded and the sets are expected to exist in "*D:\webgraph\sets*" folder while the catalogue structure of output files is as follows:

- The root folder is located in "*D:\webgraph\stripes*".

- For each *B* there exists a separate folder of name equal to *B*.

- In each of these folders for each *K* there exists a separate folder of name equal to K.

For the testing purposes the values of *B* equal to 2048, 1024, 512, 256 and 128 were provided while values 8, 16, 32, 64, 128 form chosen values of *K*. That results in 25 folders that have to be present in order for the application to work, since there are no mechanisms to check file existence and the application will crash in such condition. It was a conscious decision since the program is not intended to be user friendly but to provide proof of concept as fast as possible, so some assumptions were made beforehand in order to reduce testing time.

If one needs to change catalogues, names of sets, the number of sets, *B* or *K* parameters it is necessary to recompile the main function. Thus the tests were made in the main application for speed purposes, since invoking application frequently can cause memory swaps and other slowdowns which was unwanted behaviour.

## 5.1. Uncompressed tile format

Since there are two ways for the contents of a tile to be renumbered (horizontally and vertically) it is convenient to store both numerations at once and while introducing new content (adding link) during line processing both values are calculated at once. In such a way it is not necessary to make a transpose of one list of residues after tile creation, but it is sufficient to go through all elements of a tile only once. One may notice that for horizontal numbering there would be created a list in increasing order (since rows of input text file are processed from left to right and from top to bottom, the resulting number, which is number of the column in a tile increased by multiplication of number of the row in a tile and tile size, will be increasing) and for vertical numbering this will not be the case (since the resulting number, which is number of the row in a tile increased by multiplication of number of the column and tile size, can be greater or smaller). Due to this fact the former values can be stored in an unsorted array while the latter need some sorted structure. Here the STL vector and set were used respectively. Another observation was made that since both structures will be differentially encoded later, the horizontal one can be calculated on the fly (since next values can be greater only).

As mentioned in the previous chapter the stripes are stored using bit arrays which is implemented using unsigned char arrays of size equal to number of required bits divided by 8 (and then each bit is accessed using bitwise logic). The bits are set during insertion of a link into the tile, which is presented in code snippet 3.

```cpp
void CUncompressedTile::addLink(int x, int y) {

        xDiffs.push_back((y << logTileSize) + x - last);
        last = (y << logTileSize) + x;

        yDiffs.insert((x << logTileSize) + y);

        int tmpX = (x * stripSize) >> logTileSize; // mark columns (that they contain something)
        xStrip[tmpX >> 3] |= 1 << (tmpX & 7);

        int tmpY = (y * stripSize) >> logTileSize; // mark rows
        yStrip[tmpY >> 3] |= 1 << (tmpY & 7);
}
```

*Code snippet 3. Adding link into the tile.*

Since tile size $B$ and stripes amount $K$ are indispensable to calculate all values it is necessary to specify them after tile creation (the tiles are to be constructed using array constructor, so it would be impossible to pass any arguments to constructors).

Compression of a tile is performed when all values are known (after the last possible link was added). It consists of byte encoding horizontal differences (and putting output to the first buffer), zipping the output of the just finished byte encoding using Deflate (and putting output to the second buffer), differential encoding and then byte encoding of vertically reordered edges (and putting output to the third buffer) and finally zipping the output of the previous stage using Deflate (and putting output to the fourth buffer). For byte encoding the assumption was made that each difference can be expressed in maximum 3 bytes since for the largest tile size $B = 2048$ it would require the highest jump of 2048*2048 - 1 = 4194303, which is almost $2^{22}$ and this is exactly maximum number of values 3 bytes can contain (assuming that two bits are left for indication whether the value is coded using 1, 2 or 3 bytes). Hence the length of the buffer should be at least 3 times greater than the number of residues plus additional byte for encoding zero at the end of residues list. The idea behind byte encoding is to represent the differences in as little bytes as possible. Since the differences are usually small (due to the locality of a web graph), they are coded using 1 byte, which is much less than using typical 4 byte integer representation. The compressing function is shown in code snippet 4.

```cpp
int CUncompressedTile::compress(int xTile, int yTile, unsigned char *buffer, int &startBuf) {
    // try four methods and choose the best one

    //method 1: byte encode x diffs:
    unsigned char *buf1 = new unsigned char[xDiffs.size() * 3 + 1];
    int startBuf1 = 0;
    for(auto it=xDiffs.begin(); it!=xDiffs.end(); it++)
        byteEncode(*it, buf1, startBuf1);
    byteEncode(0, buf1, startBuf1);

    //method 2: zip byte encode x diffs
    int startBuf2 = (int)xDiffs.size() * 5 + 1;
    unsigned char *buf2 = new unsigned char[startBuf2];
    deflateBytes(buf1, startBuf1, buf2, startBuf2);

    //method 3: byte encode y diffs (x transposed)
    unsigned char *buf3 = new unsigned char[yDiffs.size() * 3 + 1];
    int startBuf3 = 0;
    last = -1;
    for(auto it=yDiffs.begin(); it!=yDiffs.end(); it++) {
        byteEncode((*it)-last, buf3, startBuf3);
        last = *it;
    }
    byteEncode(0, buf3, startBuf3);

    //method 4: zip byte encode y diffs (x transposed)
    int startBuf4 = (int)yDiffs.size() * 5 + 1;
    unsigned char *buf4 = new unsigned char[startBuf4];
    deflateBytes(buf3, startBuf3, buf4, startBuf4);

    //choose method
    int ret = startBuf;
    if(startBuf1 <= startBuf2)
        if(startBuf3 <= startBuf4)
            if(startBuf1 <= startBuf3) {
                // choose 1
                memcpy(buffer + startBuf, buf1, startBuf1);
                startBuf += startBuf1;
            } else {
                // choose 3
                memcpy(buffer + startBuf, buf3, startBuf3);
                startBuf += startBuf3;
                ret |= 2 << 30; //(2 << 30);
            }
        else
            if(startBuf1 <= startBuf4) {
                // choose 1
                memcpy(buffer + startBuf, buf1, startBuf1);
                startBuf += startBuf1;
            } else {
                // choose 4
                memcpy(buffer + startBuf, buf4, startBuf4);
                startBuf += startBuf4;
                ret |= 3 << 30;
            }
    else
        if(startBuf3 <= startBuf4)
            if(startBuf2 <= startBuf3) {
                // choose 2
                memcpy(buffer + startBuf, buf2, startBuf2);
                startBuf += startBuf2;
                ret |= 1 << 30;
            } else {
                // choose 3
                memcpy(buffer + startBuf, buf3, startBuf3);
                startBuf += startBuf3;
                ret |= 2 << 30;
            }
        else
            if(startBuf2 <= startBuf4) {
                // choose 2
                memcpy(buffer + startBuf, buf2, startBuf2);
                startBuf += startBuf2;
                ret |= 1 << 30;
```

```
                } else {
                        // choose 4
                        memcpy(buffer + startBuf, buf4, startBuf4);
                        startBuf += startBuf4;
                        ret |= 3 << 30;
                }

        delete[] buf1;
        delete[] buf2;
        delete[] buf3;
        delete[] buf4;

        // save xStrip and yStrip as well
        int bytesPerStrip = (stripSize+7) >> 3;
        memcpy(buffer + startBuf, xStrip, bytesPerStrip);
        startBuf += bytesPerStrip;
        memcpy(buffer + startBuf, yStrip, bytesPerStrip);
        startBuf += bytesPerStrip;

        // and x and y of a tile
        buffer[startBuf++] = xTile & 255;
        buffer[startBuf++] = (xTile >> 8) & 255;
        buffer[startBuf++] = xTile >> 16;

        buffer[startBuf++] = yTile & 255;
        buffer[startBuf++] = (yTile >> 8) & 255;
        buffer[startBuf++] = yTile >> 16;

        return ret;
}
```

*Code snippet 4. The tile compression.*

Deflate algorithm mostly outputs less bytes than it is given, but there exist extreme cases, so for safety reasons the second buffer has its size doubled compared to the first one. The third and fourth buffers are of the same sizes as first and second respectively for the same reasons. After all processing is done the method that produced the smallest output is chosen and the results are being concatenated to the output buffer. If more than one method produced the smallest output, the quickest one to calculate is chosen.

## 5.2. Compressed tile format

A compressed tile object is used to extract information about the edges of web graph. The tile size $B$, stripes amount $K$, input buffer and output array has to be defined before extracting any data. What is more it is needed to specify the row in a tile before extracting successors or the column in a tile before extracting predecessors. In this way all the redundant data are omitted between consecutive method calls (there are many tiles in a row or column to be decompressed while retrieving successor or predecessor lists).

The decompression itself involves checking the stripe flag in order to test whether the tile contains any edge in the specified row or column. This is the main upgrade of the simple 2D method described in the previous chapter. If there exists at least one edge, the whole tile is then processed. First it is checked which packing type was used during compression stage (this information is stored in two most significant bits of the offset). Then appropriate decoding follows. If the tile was zipped, it is uncompressed and then processed according to the need.

If there is a need for successors, then the horizontal representation is differentially decoded and checked to contain values from the proper range. If the values go beyond the range, the analysis stops. If the tile is coded using vertical representation, it has to be differentially decoded, transposed and then checked for the specific range, but since it is transposed, we have to decode all values, because they are not monotonous. This explains the line count of code snippet 5.

For the list of predecessors it is the other way round which means that the horizontal representation has to transposed and its processing cannot be ended before the last value is obtained. All values that belong to the range are then recalculated, because only their position in a tile is known, so it is necessary to add tile offset in an adjacency matrix to obtain correct node number. Finally they are stored in an array that was forwarded at the beginning of retrieval.

### 5.3. Compressing the graph

As stated at the beginning of this chapter the input file is buffered for efficiency. The buffer is set to 32 MB, but can be adjusted. There is a search performed to count the lines (new line character occurrences) each time the buffer is read. Then all bytes up to the last new line character are being processed. After it is finished, the rest of buffer is being moved at the beginning and the next chunk is being added from file. The operation repeats until all the bytes have been read and file is closed.

Each line is processed character by character. If it is a digit, temporal value is being multiplied by ten and added with the numerical value of character (decreased by 48 because of ASCII coding) in order to obtain the same successor number as written in file. If the character being read is a space, that adds the link to the corresponding tile. If the character is a new line the processing of the line ends. It is shown in code snippet 2.

In this idea the whole adjacency matrix is being divided into squares, so in one row there exist many tiles (equal to number of nodes divided by tile size $B$). If we wanted to represent all tiles at once and then save only those that are empty, we would need unimaginably huge amount of memory (for Arabic-2005 set it would require storing $(22\,744\,080\,/\,128)^2 = 31\,573\,380\,721$ tiles for $B = 128$, that is over 30 GB assuming 1 B per tile which is unrealistic). However, thanks to the fact that we are processing input row by row it is sufficient to store only one row of tiles at a time and then compress only non-empty ones.

While being compressed, the row and the column of the tile in the tiles matrix are added to the tile compressed representation in order to be able to recover node numbers while retrieving lists. Additionally the offset is being registered in a vector of horizontal offsets (*xOffsetsVector*) and added to corresponding vector of vertical vectors (*yOffsetsVectors*). The reason for introducing vertical vectors is that the amount of non-empty tiles in each column is not known before the whole input is processed and it is impossible to assign the first offset of a tile in the following column without knowing this number. So those offsets have to be stored in different vectors and then merged after the whole input is processed.

In order to be able to recover whole line we need to know which tile is first in a row or column of our interest. For that purpose while processing row of tiles the offset of the first one is being stored in an array of first tile offsets in a row (*xFirstOffsets*). For the same reason as we needed vertical vectors it is impossible to determine the vertical first offsets at this moment, so they are being calculated after the whole input processing is done and stored in and array of first tile offsets in a column (*yFirstOffsets*). Both arrays have lengths equal to number of nodes divided by tile size B, which is known before the processing, thus can be created by then.

After the input has been processed the last thing that has to be done is to calculate vertical offsets, fill array of first tile offsets in a column, create and initialize one tile that is used for decompression of a graph and create output results array of retrieving successors or predecessors (in order to increase speed).

### 5.4. Retrieving the list

Since both successor and predecessor retrieval expose similarities (they are symmetric) the process of extracting successor list is going to be shown thoroughly. First we need to calculate the row of tiles that cover the row in adjacency matrix, which is simply node number divided by tile size. Then the offset of the leftmost tile is taken from the array of first offsets and all consecutive tiles are being processed until the first tile of the next row is encountered. Tile offsets contain the information about the type of encoding used for tile compression stored in their two most significant bits, so if one is confused about uncommon values of offsets, it is due to the fact that two bits have to be truncated in order to obtain the real offset value. Tile decompression was discussed in detail in section 5.2 "Compressed tile format".

```cpp
void CCompressedTile::decompressX(int offset, int nextOffset) {
        offsetVal = offset & 0x3FFFFFFF;
        nextOffsetVal = nextOffset & 0x3FFFFFFF;
        length = nextOffsetVal - offsetVal;
        x = (buffer[nextOffsetVal-6] + ((int)buffer[nextOffsetVal-5] << 8) +
((int)buffer[nextOffsetVal-4] << 16)) << logTileSize; // copy x offset and multiply by tilesize
        if((buffer[nextOffsetVal-yStripOffset] & tmpYmask) != 0) { // if there is something
                if((offset & 0x80000000) == 0) { // normal, not transposed
                        if((offset & 0x40000000) == 0) { // no zip
                                last = -1;
                                tmp = nextInt(buffer, offsetVal);
                                while(tmp != 0) {
                                        last += tmp;
                                        if(last >= interestedFrom)
                                                if(last < interestedTo)
                                                        outArr[len++] = x + last -
interestedFrom;
                                                else
                                                        break;
                                        tmp = nextInt(buffer, offsetVal);
                                }
                        } else { // zipped
                                tmpLen = defaultTmpLen;
                                inflateBytes(buffer+offsetVal, length-lengthOffset, tmpB,
tmpLen);
                                tmpStart = 0;
                                last = -1;
                                tmp = nextInt(tmpB, tmpStart);
                                while(tmp != 0) {
                                        last += tmp;
                                        if(last >= interestedFrom)
                                                if(last < interestedTo)
                                                        outArr[len++] = x + last -
interestedFrom;
                                                else
                                                        break;
                                        tmp = nextInt(tmpB, tmpStart);
                                }
                        }
                } else { // transposed
                        if((offset & 0x40000000) == 0) { // no zip
                                last = -1;
                                tmp = nextInt(buffer, offsetVal);
                                while(tmp != 0) {
                                        last += tmp;
                                        lastX = last >> logTileSize; // transpose
                                        lastY = last - (lastX << logTileSize);
                                        if(lastY == y)
                                                outArr[len++] = x + lastX;
                                        tmp = nextInt(buffer, offsetVal);
                                }
                        } else { // zipped
                                tmpLen = defaultTmpLen;
                                inflateBytes(buffer+offsetVal, length-lengthOffset, tmpB,
tmpLen);
                                tmpStart = 0;
                                last = -1;
```

```
tmp = nextInt(tmpB, tmpStart);
while(tmp != 0) {
        last += tmp;
        lastX = last >> logTileSize; // transpose
        lastY = last - (lastX << logTileSize);
        if(lastY == y)
                outArr[len++] = x + lastX;
        tmp = nextInt(tmpB, tmpStart);
    }
            }
        }
    }
}
```

*Code snippet 5. The tile decompression.*

# 6. RESULTS

## 6.1.Sets used

As mentioned in the introduction, the sets most commonly used for testing web graph compression algorithms are EU-2005, Indochina-2004 and UK-2002 gathered and prepared by UbiCrawler[26] of LAW. In order to improve tests quality another set was chosen, Arabic-2005, which has twice as many arcs than UK-2002 do and only 23% more nodes, showing different characteristics from other sets, therefore providing more reliable test results. Table 2 shows some characteristics of the sets used.

| set | nodes | links |
|---|---|---|
| EU-2005 | 862 664 | 19 235 140 |
| Indochina-2004 | 7 414 866 | 194 109 311 |
| UK-2002 | 18 520 486 | 298 113 762 |
| Arabic-2005 | 22 744 080 | 639 999 458 |

*Table 2. Data sets characteristics.*

The sets were downloaded from LAW website and converted into text format using Java program used for LM algorithm testing in [10]. In order to test access to predecessors the transposed graph representations had to be created for 1D methods. For the testing purpose 100 000 randomly selected nodes were chosen for each set separately in order to compare all methods using same sample. This methodology was first introduced in [12]. Text web graph files characteristics are presented in table 3.

| set | bytes |
|---|---|
| EU-2005 | 133 651 465 |
| EU-2005t | 133 575 864 |
| Indochina-2004 | 1 494 552 255 |
| Indochina-2004t | 1 492 772 153 |
| UK-2002 | 2 529 010 522 |
| UK-2002t | 2 527 404 371 |
| Arabic-2005 | 5 509 322 539 |
| Arabic-2005t | 5 488 874 469 |

*Table 3. Text web graph file sizes.*

---

[26] http://vigna.dsi.unimi.it/ftp/papers/UbiCrawler.pdf [May, 2012]

The BV framework version 3.0.3 was downloaded from the website[27] and run on Java Development Kit version 1.7.0_03 x64. The $k^2$partitioned sources were downloaded from the website[28] and compiled using Microsoft Visual Studio 2010 using Release x64 configuration. The sources needed to be modified slightly. The main differences were noticed in names of I/O functions and sizes of unsigned long and unsigned long long.

LM, 2D and 2D-stripes programs were written from scratch and compiled using the same configuration. All tests were performed on two machines: PC with Intel Core 2 Quad Q6600 2.7 GHz 2x4 MB L2 processor, 8 (4+4) GB DDR3-1200 Corsair memory and Seagate Barracuda 7200.10 320 GB, 7200 rpm, SATA II hard drive and a notebook Samsung 300V5A-S05 with Intel Core i5-2450M 2.9 GHz 2x256 KB L2, 3 MB L3 processor, 6 (2+4) DDR3-1333 Samsung memory and Samsung SpinPoint M8 1000 GB, 5400 rpm, SATA II hard drive. The former is noted as *PC* in tests and the latter is labelled *Notebook*.

Both machines run Microsoft Windows 7 SP1 x64 operating systems with PC running Ultimate version and Notebook running Home Premium. Both computers had a separate partition created which was at least 40% free during the tests.

All the compressions were performed only once so sometimes compression times may be affected by memory swapping or buffering time. However, random access tests were performed several times in a row: 10 times for BV in the same execution, 10 times for LM C++ version in the same execution, 4 times executed LM in Java, 4 times for 2D and 2D stripes both in one execution and $k^2$partitioned executed 3 times in a row. For all tests the average time was measured apart from $k^2$partitioned, where the best time was taken in order to reduce the gap between our solutions.

The output of LM algorithms was then transformed back to text form and verified with the original file in order to assure correctness.

---

[27] http://webgraph.dsi.unimi.it/ [May, 2012]
[28] http://webgraphs.recoded.cl/index.php?section=k2part [May, 2012]

### 6.2. Compression

Table 4 presents the compression times for two LM implementations.

| C++ (PC) | EU | EU-t | Indo | Indo-t | UK | UK-t | Arabic | Arabic-t |
|---|---|---|---|---|---|---|---|---|
| 8 | 6.8 | 8.7 | 52.7 | 66.6 | 184.3 | 164.5 | 228.4 | 404.3 |
| 16 | 6.2 | 8.1 | 56.6 | 64.9 | 103.9 | 120.7 | 207.9 | 276.7 |
| 32 | 6.2 | 9.2 | 54.7 | 66.8 | 101.2 | 153.9 | 212.3 | 322.0 |
| 64 | 8.9 | 12.2 | 64.0 | 80.7 | 128.5 | 178.1 | 322.2 | 454.6 |
| 128 | 13.6 | 17.7 | 88.1 | 111.2 | 213.1 | 301.5 | 491.0 | 749.8 |
| C++ (Notebook) | EU | EU-t | Indo | Indo-t | UK | UK-t | Arabic | Arabic-t |
| 8 | 4.7 | 8.8 | 56.0 | 60.8 | 106.6 | 133.5 | 267.2 | 295.5 |
| 16 | 5.7 | 7.6 | 50.1 | 58.6 | 93.5 | 112.2 | 257.3 | 319.8 |
| 32 | 6.0 | 9.5 | 49.8 | 64.5 | 98.4 | 138.3 | 268.9 | 393.5 |
| 64 | 9.6 | 13.0 | 63.2 | 74.9 | 122.3 | 182.1 | 338.4 | 516.0 |
| 128 | 12.9 | 16.8 | 78.9 | 97.6 | 205.3 | 301.1 | 481.9 | 741.8 |
| Java (PC) | EU | EU-t | Indo | Indo-t | UK | UK-t | Arabic | Arabic-t |
| 8 | 9.7 | 14.6 | 82.2 | 96.7 | 143.2 | 167.1 | 285.3 | 367.0 |
| 16 | 9.9 | 14.0 | 78.1 | 97.5 | 132.3 | 163.1 | 269.1 | 373.4 |
| 32 | 11.0 | 17.5 | 82.5 | 112.3 | 173.7 | 181.5 | 287.4 | 472.3 |
| 64 | 16.3 | 24.7 | 97.2 | 141.0 | 174.2 | 247.8 | 399.8 | 707.2 |
| 128 | 27.0 | 40.1 | 146.9 | 219.6 | 282.3 | 465.8 | 607.5 | 987.3 |
| Java (Notebook) | EU | EU-t | Indo | Indo-t | UK | UK-t | Arabic | Arabic-t |
| 8 | 7.1 | 9.2 | 56.0 | 68.0 | 97.9 | 126.2 | 212.4 | 278.7 |
| 16 | 7.3 | 11.2 | 54.5 | 70.5 | 91.7 | 129.1 | 206.4 | 291.1 |
| 32 | 8.2 | 14.1 | 59.3 | 79.3 | 95.6 | 137.6 | 224.8 | 395.8 |
| 64 | 13.4 | 20.2 | 75.9 | 115.4 | 134.8 | 197.8 | 336.1 | 536.9 |
| 128 | 21.2 | 30.6 | 123.0 | 165.0 | 226.8 | 328.7 | 526.3 | 758.6 |

*Table 4. LM compression time.*

First of all let us present the result of different approaches for LM compression. The original LM compressor written in Java has performed all operations on both computers in 14568.986 s while its C++ counterpart finished the task in 12245.623 s. The speed increase is 19 % on average, however, one may notice that there are cases when C++ variant is slower (LM-8 for UK-2002 or Arabic-2005t on PC) than Java equivalent. The reason for such behaviour was discussed earlier. Anyhow the tests show that rewriting code in C++ was worth the effort. It is especially visible when compared to its biggest opponent, BV scheme, whose compression times are presented in table 5.

| BV (PC) | EU | EU-t | Indo | Indo-t | UK | UK-t | Arabic | Arabic-t |
|---|---|---|---|---|---|---|---|---|
| | 10.1 | 10.8 | 73.4 | 94.9 | 136.8 | 152.2 | 275.0 | 329.8 |
| BV (Notebook) | EU | EU-t | Indo | Indo-t | UK | UK-t | Arabic | Arabic-t |
| | 7.3 | 7.9 | 66.6 | 66.0 | 110.9 | 129.7 | 234.2 | 250.3 |

*Table 5. BV compression time.*

LM Java variant is slower at compression in every possible case comparing to BV, yet C++ version is faster with LM-16 variant and mostly in LM-32. It beats BV scheme with respect to compression time while achieving satisfactory or even better ratios, which will be covered later.

Let us compare compression times of 2D techniques which are shown in table 6.

| 2Dstripes (PC) | | EU-2005 | Indochina-2004 | UK-2002 | Arabic-2005 |
|---|---|---|---|---|---|
| B | K | | | | |
| 2048 | 8 | 13.0 | 196.1 | 387.7 | 707.4 |
| | 16 | 14.1 | 158.3 | 361.5 | 613.5 |
| | 32 | 13.2 | 161.2 | 324.9 | 638.4 |
| | 64 | 13.6 | 169.5 | 338.3 | 569.6 |
| | 128 | 12.6 | 162.7 | 322.2 | 604.4 |
| 1024 | 8 | 11.8 | 145.6 | 272.8 | 639.7 |
| | 16 | 11.6 | 145.5 | 281.6 | 630.4 |
| | 32 | 12.5 | 143.1 | 256.2 | 618.4 |
| | 64 | 12.0 | 143.8 | 261.7 | 613.1 |
| | 128 | 12.0 | 144.0 | 276.2 | 628.9 |
| 512 | 8 | 11.2 | 132.3 | 276.0 | 603.2 |
| | 16 | 11.2 | 135.8 | 279.8 | 620.5 |
| | 32 | 11.5 | 135.3 | 230.0 | 642.6 |
| | 64 | 11.6 | 143.8 | 239.6 | 638.5 |
| | 128 | 11.6 | 144.6 | 252.4 | 551.4 |
| 256 | 8 | 11.9 | 155.4 | 348.4 | 576.7 |
| | 16 | 11.8 | 139.4 | 374.9 | 568.1 |
| | 32 | 11.8 | 140.5 | 371.9 | 569.7 |
| | 64 | 11.8 | 143.1 | 374.4 | 614.5 |
| | 128 | 12.2 | 144.7 | 378.9 | 625.2 |
| 128 | 8 | 14.0 | 224.2 | 704.0 | 1091.8 |
| | 16 | 14.3 | 216.0 | 830.7 | 1069.0 |
| | 32 | 14.7 | 174.2 | 679.8 | 1049.1 |
| | 64 | 15.0 | 163.4 | 825.8 | 1087.8 |
| | 128 | 14.3 | 151.8 | 864.5 | 1137.2 |

| Nostripes (PC) | | EU-2005 | Indochina-2004 | UK-2002 | Arabic-2005 |
|---|---|---|---|---|---|
| | B | | | | |
| | 2048 | 12.8 | 149.2 | 322.8 | 582.6 |
| | 1024 | 9.9 | 110.9 | 210.7 | 463.4 |
| | 512 | 9.6 | 101.6 | 194.6 | 462.8 |
| | 256 | 10.0 | 100.4 | 254.9 | 507.7 |
| | 128 | 11.9 | 134.5 | 511.3 | 893.0 |
| **2Dstripes (Notebook)** | | **EU-2005** | **Indochina-2004** | **UK-2002** | **Arabic-2005** |
| **B** | **K** | | | | |
| 2048 | 8 | 9.2 | 116.1 | 326.6 | 542.0 |
| | 16 | 9.4 | 118.0 | 265.8 | 552.4 |
| | 32 | 9.7 | 119.9 | 268.4 | 557.6 |
| | 64 | 9.6 | 118.3 | 248.2 | 564.0 |
| | 128 | 9.4 | 119.7 | 266.6 | 583.8 |
| 1024 | 8 | 8.3 | 101.4 | 210.2 | 549.7 |
| | 16 | 8.4 | 101.1 | 198.7 | 572.0 |
| | 32 | 9.0 | 102.3 | 213.3 | 572.8 |
| | 64 | 8.5 | 102.5 | 199.5 | 567.4 |
| | 128 | 8.2 | 102.7 | 194.9 | 549.6 |
| 512 | 8 | 8.3 | 88.8 | 188.9 | 516.6 |
| | 16 | 7.6 | 91.3 | 172.9 | 509.6 |
| | 32 | 7.5 | 91.5 | 170.6 | 554.0 |
| | 64 | 7.5 | 91.0 | 171.0 | 592.2 |
| | 128 | 7.6 | 88.8 | 164.1 | 550.2 |
| 256 | 8 | 7.5 | 82.8 | 201.0 | 687.3 |
| | 16 | 7.5 | 80.7 | 199.3 | 556.4 |
| | 32 | 7.6 | 80.5 | 195.4 | 589.3 |
| | 64 | 7.6 | 80.3 | 192.8 | 585.9 |
| | 128 | 8.0 | 81.7 | 198.5 | 549.5 |
| 128 | 8 | 8.7 | 105.7 | 335.3 | 733.8 |
| | 16 | 8.7 | 103.3 | 331.6 | 727.8 |
| | 32 | 8.7 | 104.2 | 323.2 | 703.2 |
| | 64 | 8.7 | 104.7 | 318.9 | 717.2 |
| | 128 | 8.6 | 103.7 | 322.1 | 779.7 |
| **Nostripes (Notebook)** | | **EU-2005** | **Indochina-2004** | **UK-2002** | **Arabic-2005** |
| | B | | | | |
| | 2048 | 10.6 | 151.5 | 317.6 | 577.8 |
| | 1024 | 9.4 | 111.6 | 219.8 | 556.4 |
| | 512 | 8.7 | 96.1 | 179.1 | 499.8 |
| | 256 | 8.8 | 89.0 | 206.4 | 470.1 |
| | 128 | 10.1 | 109.2 | 321.2 | 695.3 |

*Table 6. 2D and 2D stripes compression time.*

The $k^2$partitioned compression time was omitted since simple EU-2005 compression took over half an hour and was divided into several stages using Java codes, Python script, etc. UK-2002 took almost half a day to compress and Arabic-2005 was too large for the program to finish its job in reasonable time (it needed more than 8 GB of RAM, so memory swapping occurred lengthening process couple times). For that reason only random access time was measured and without Arabic-2005.

We could logically expect that our new 2D stripes method should be slower in compression than simple 2D (called *nostripes*) due to increased structure size and processing needed by adding stripes and this is true if one observe average times of the same tile size *B* for PC. On the notebook however, this difference is much smaller and sometimes more complicated idea is even faster than the normal one (for example using set EU-2005 2D stripes outperformed 2D nostripes).

There is also another interesting observation. Together with smaller tiles the faster processing is expected as there is less amount of computation done for each tile. However, practice shows that there is a threshold beyond which the processing time increases (between $B = 512$ and $B = 256$). It was impossible to find any relation between different values of K and compression time (in fact bigger K should result in a slight slowdown as there are more bytes to be added for each tile). Still this could happen as a result of swapping or other system related actions.

In the results log file that is included on the CD attached to this thesis the compressing speed in MB/s was calculated. The average speed for 2D nostripes on a PC was 9.43 MB/s and for 2D stripes 7.69 MB/s, while on Notebook the values were 12.19 MB/s and 13.02 MB/s respectively. The reason for such contradicting results is very difficult to be found, however, it is very likely that notebook has encountered power switching activity lag or some similar power saving mechanism interruption since there were some contradicting outcomes earlier. To increase credibility of those results, the sets should be compressed a couple of times more.

### 6.3. Random access

Figures on the following pages show graphs presenting compression ratio versus random access time characteristic of all sets using different algorithms. The results of retrieving predecessor list for 2D methods queries are presented on transposed graphs for simplicity.

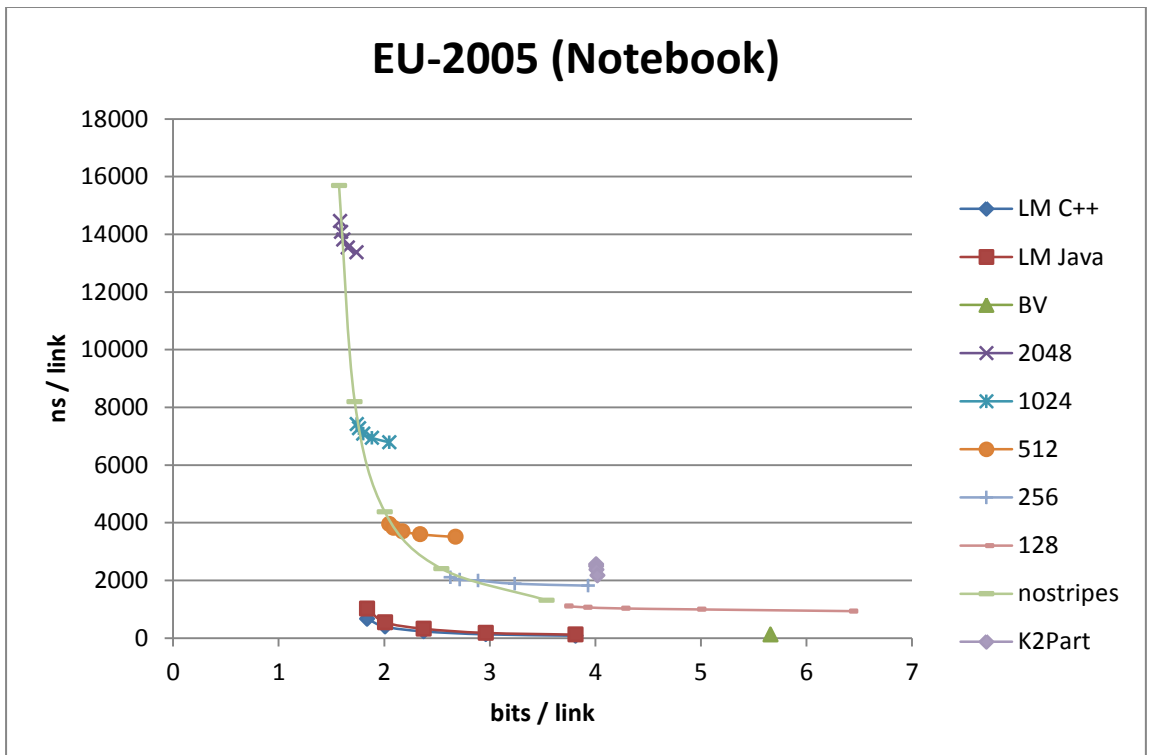

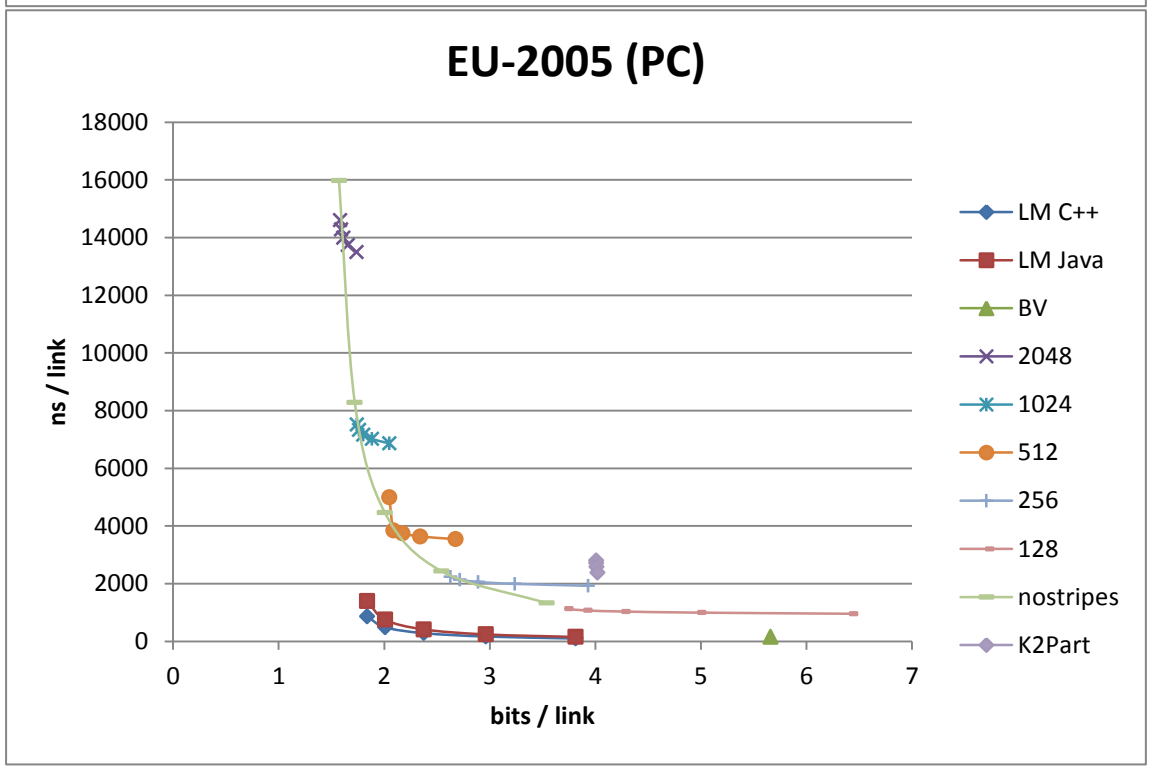

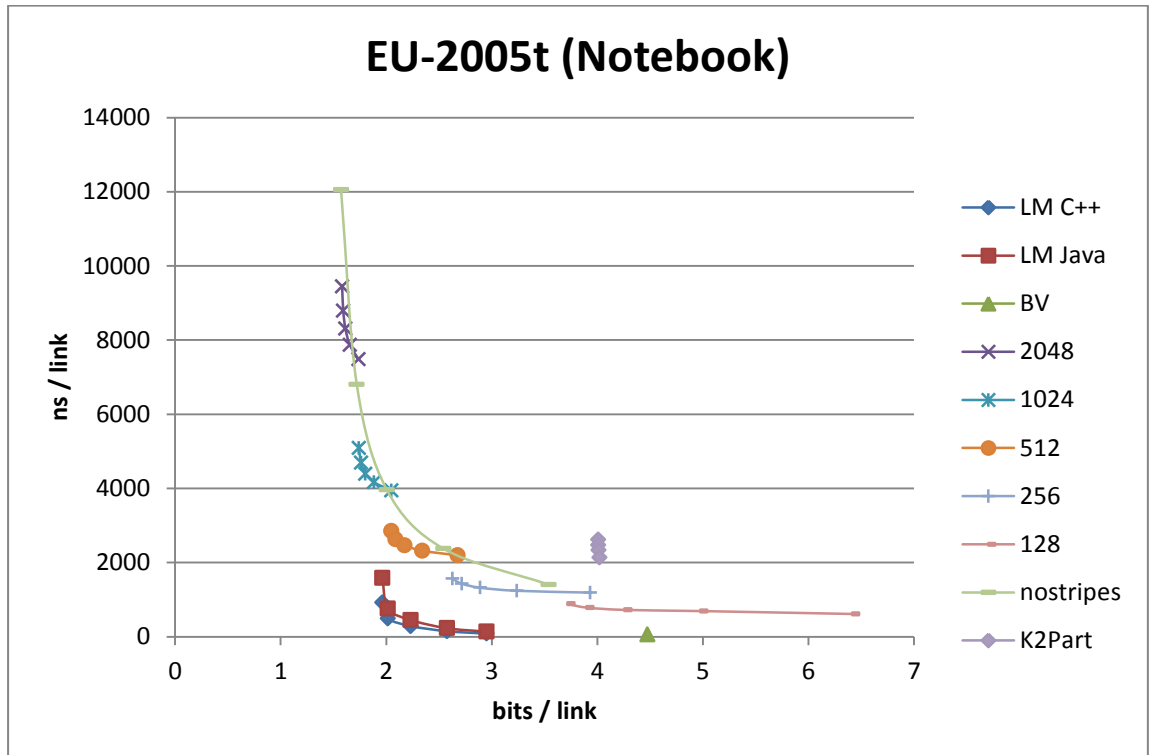

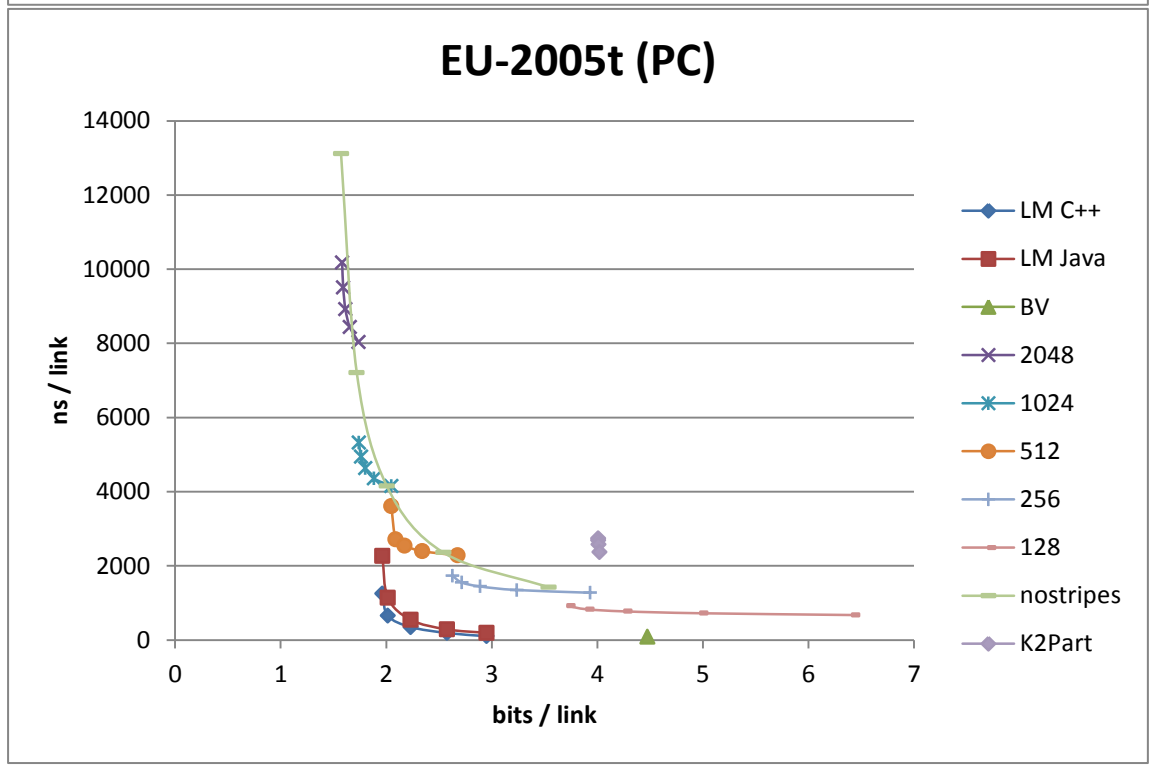

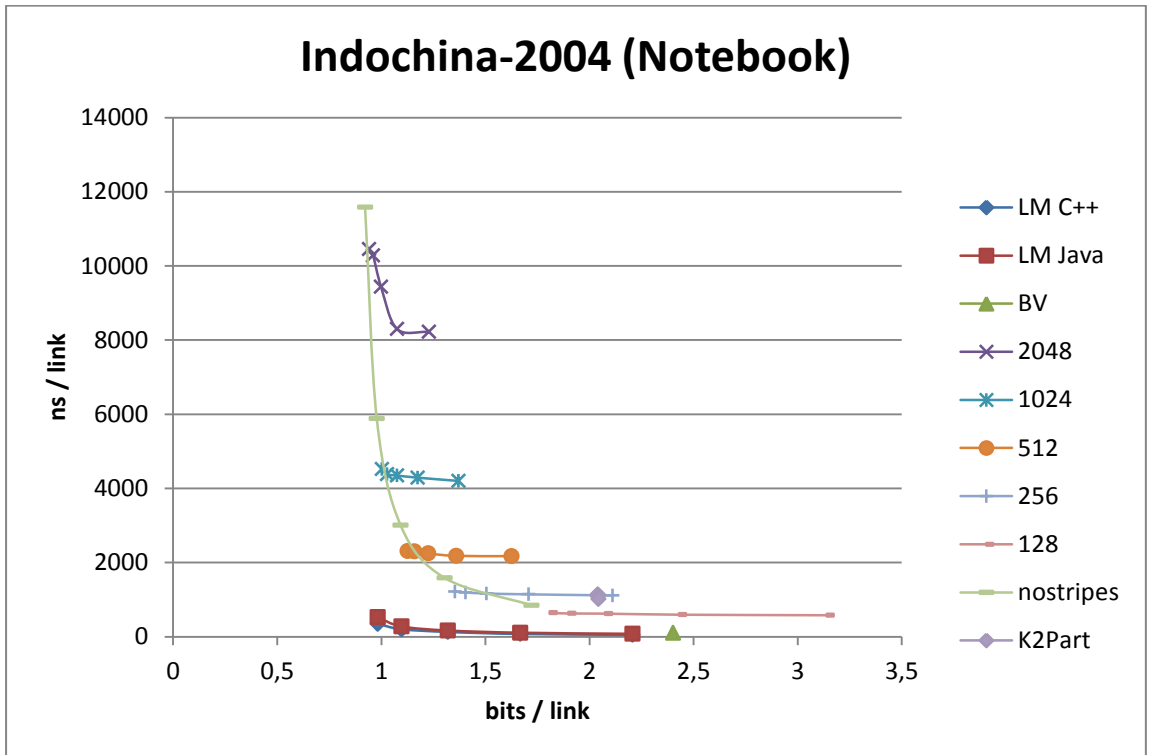

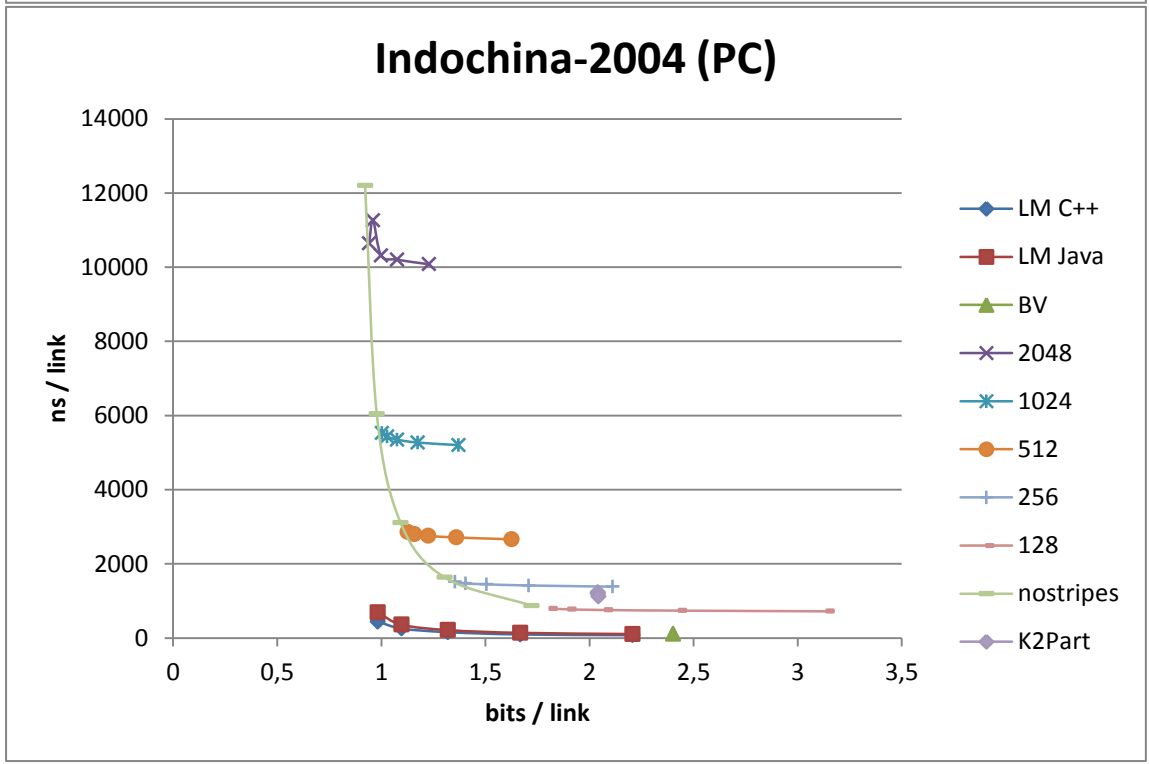

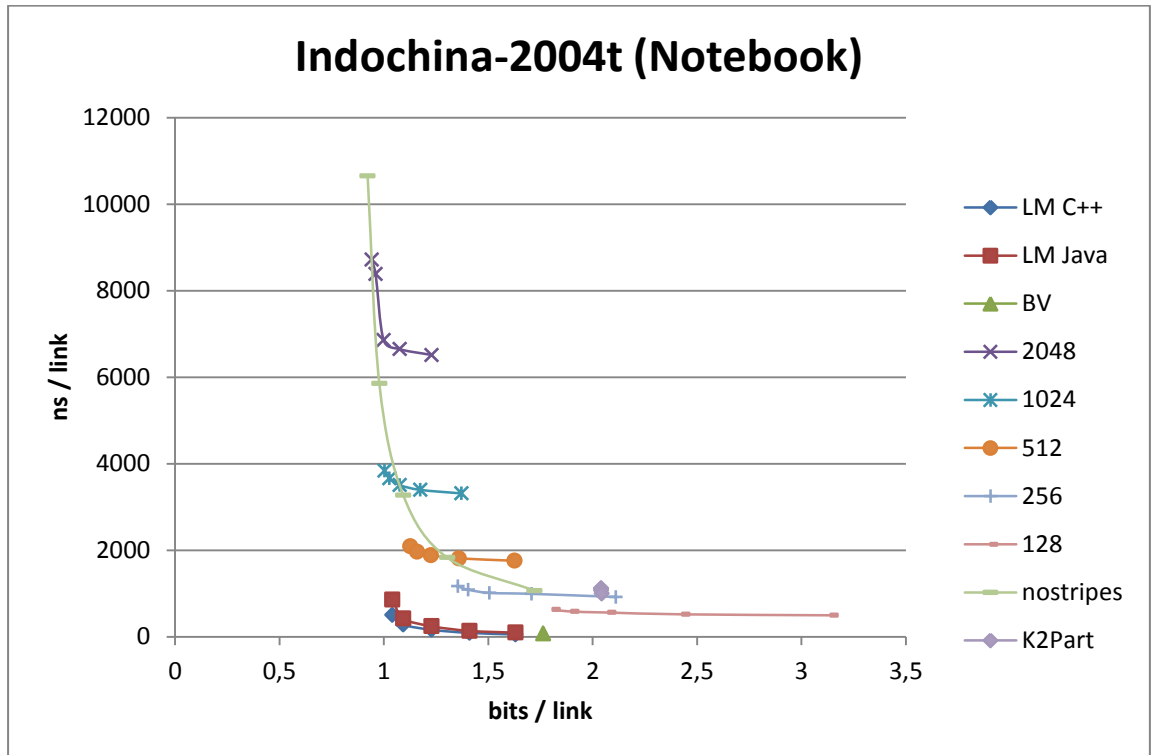

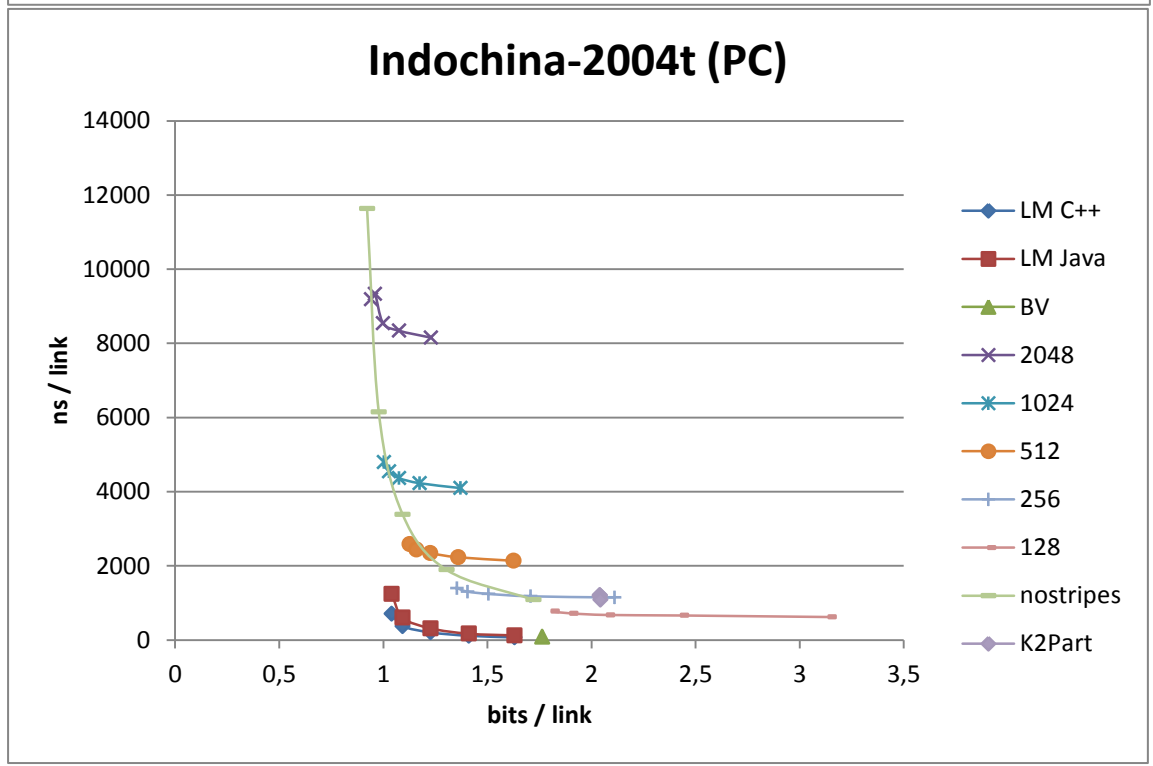

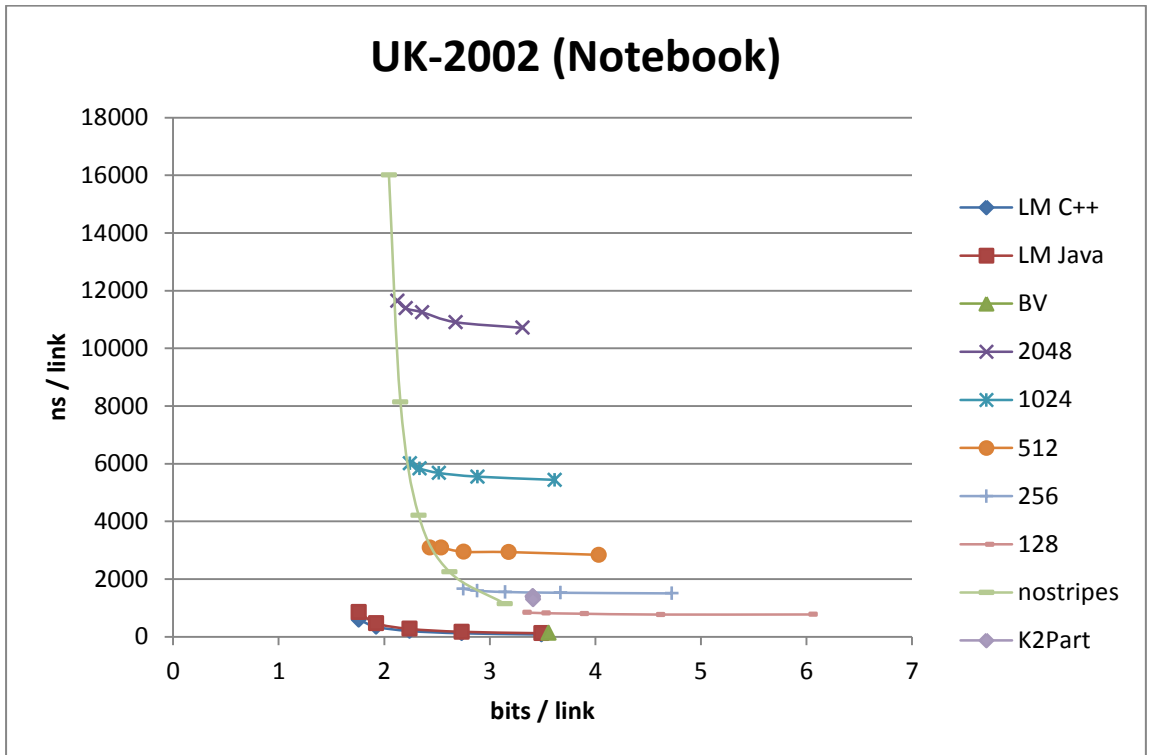

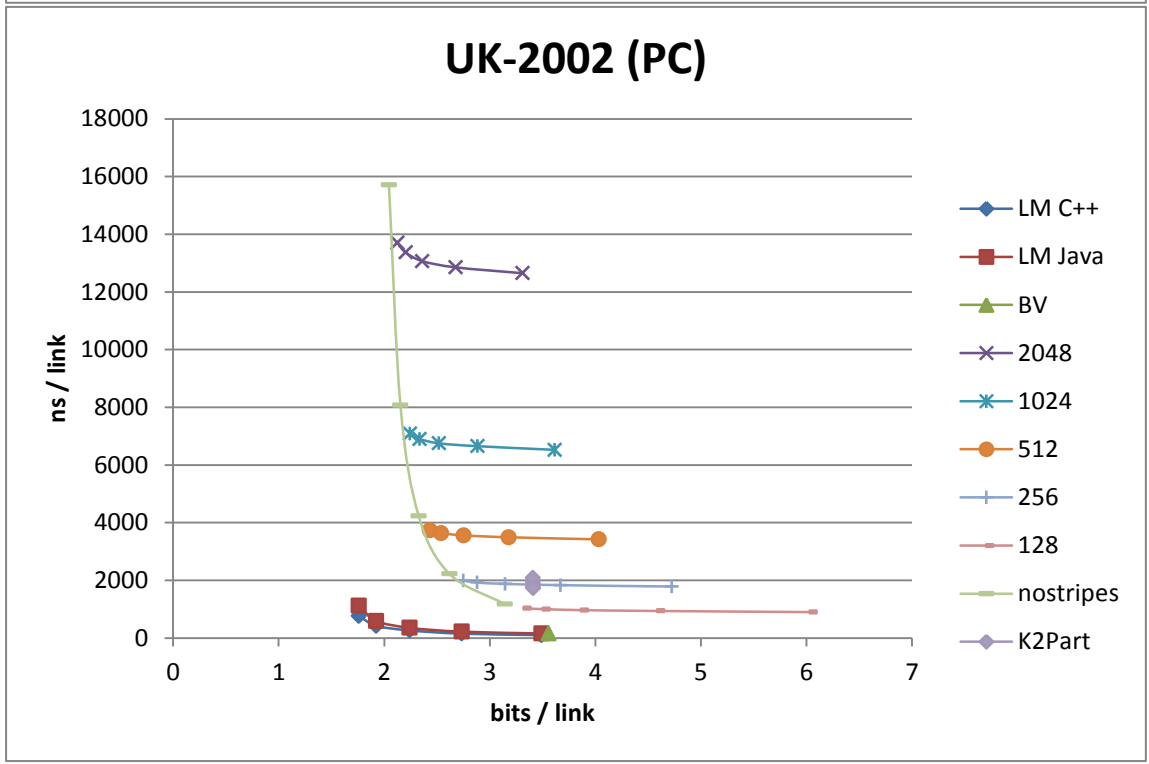

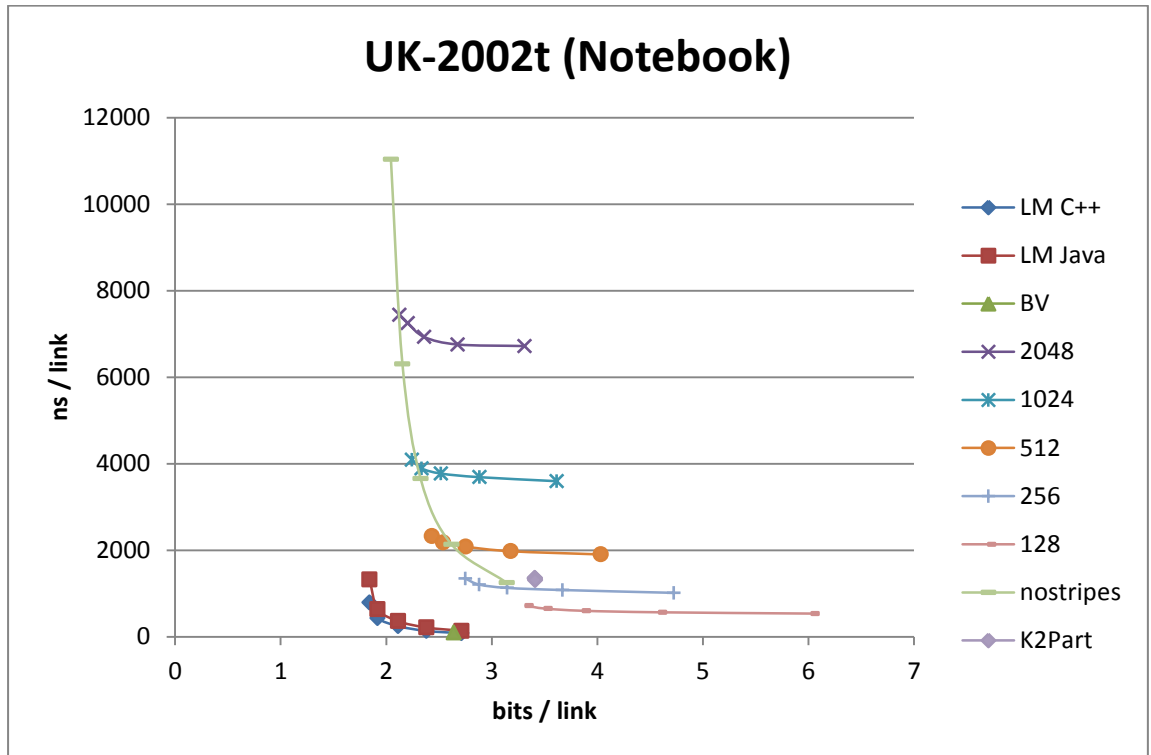

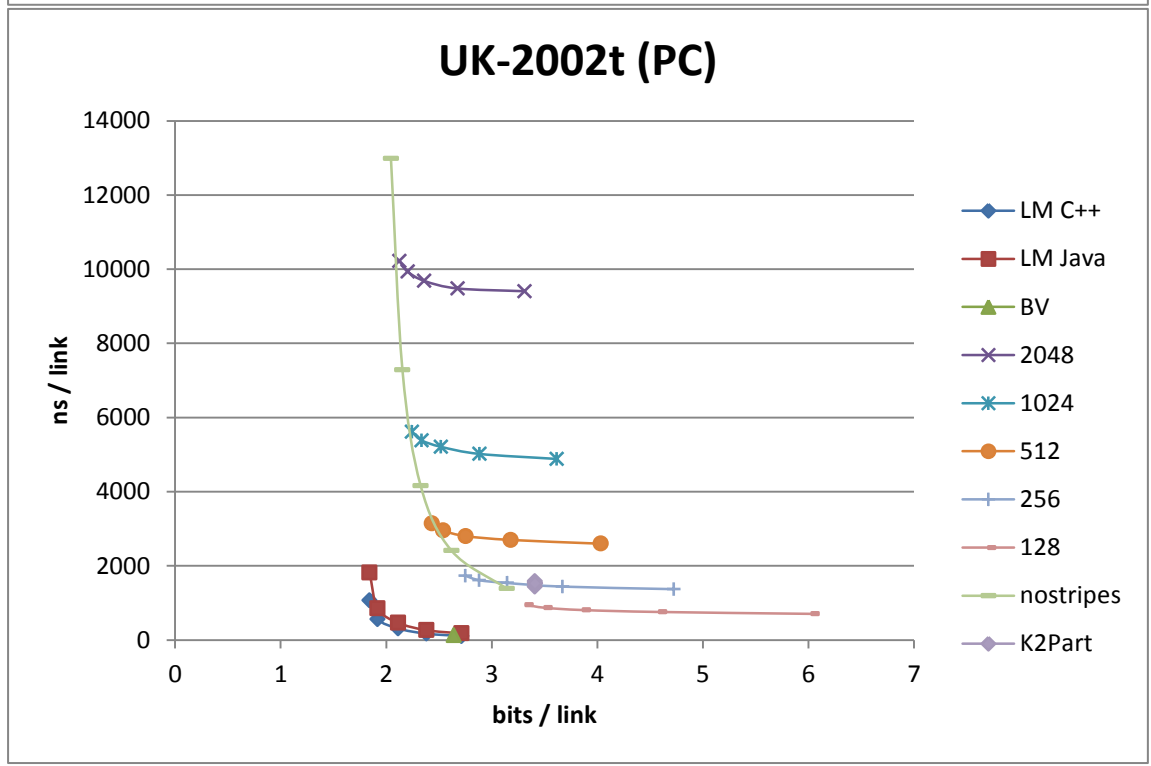

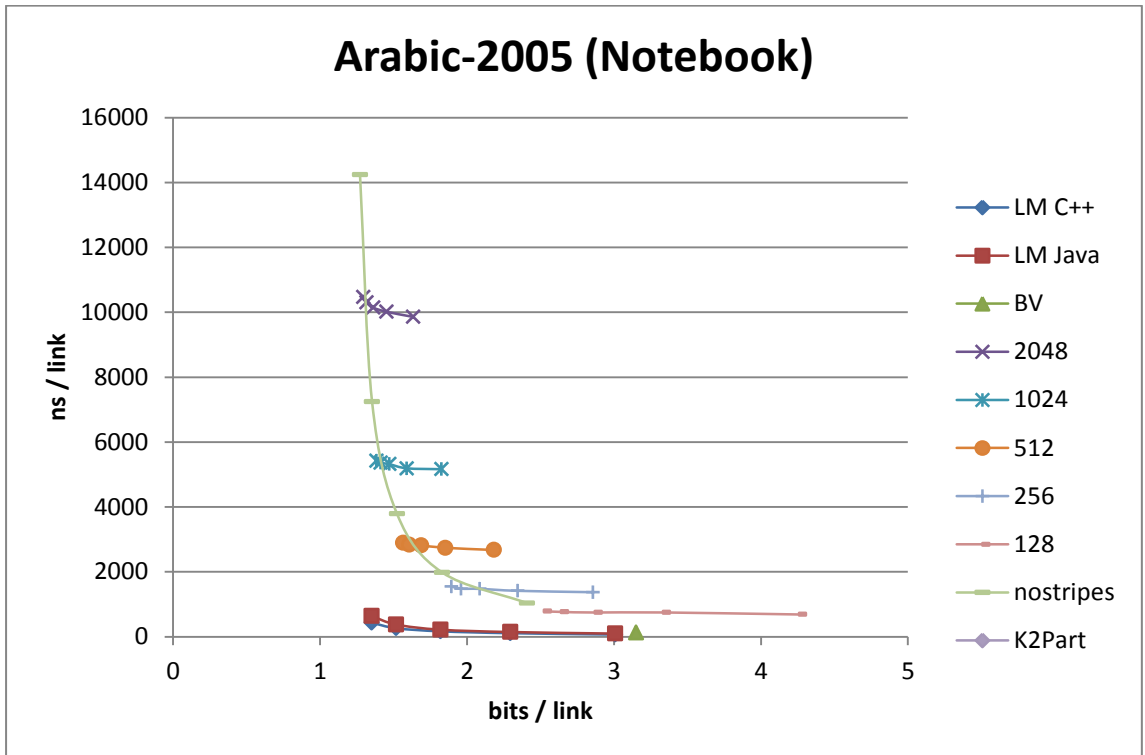

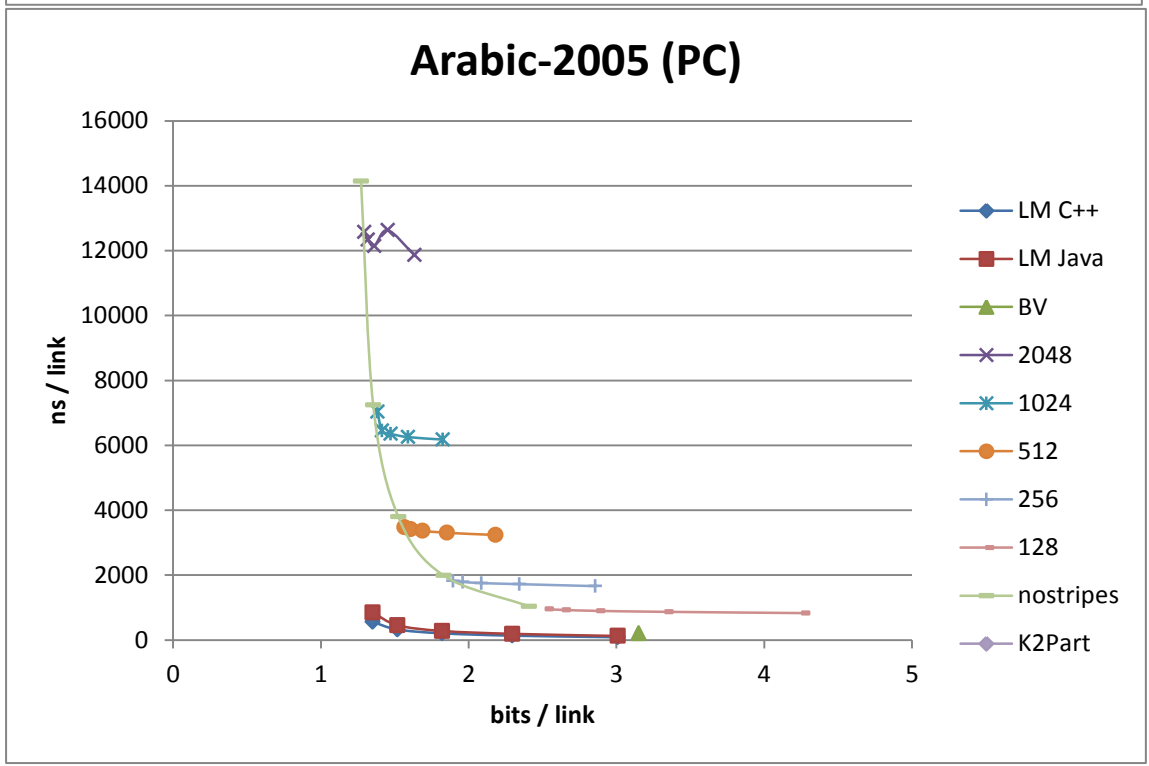

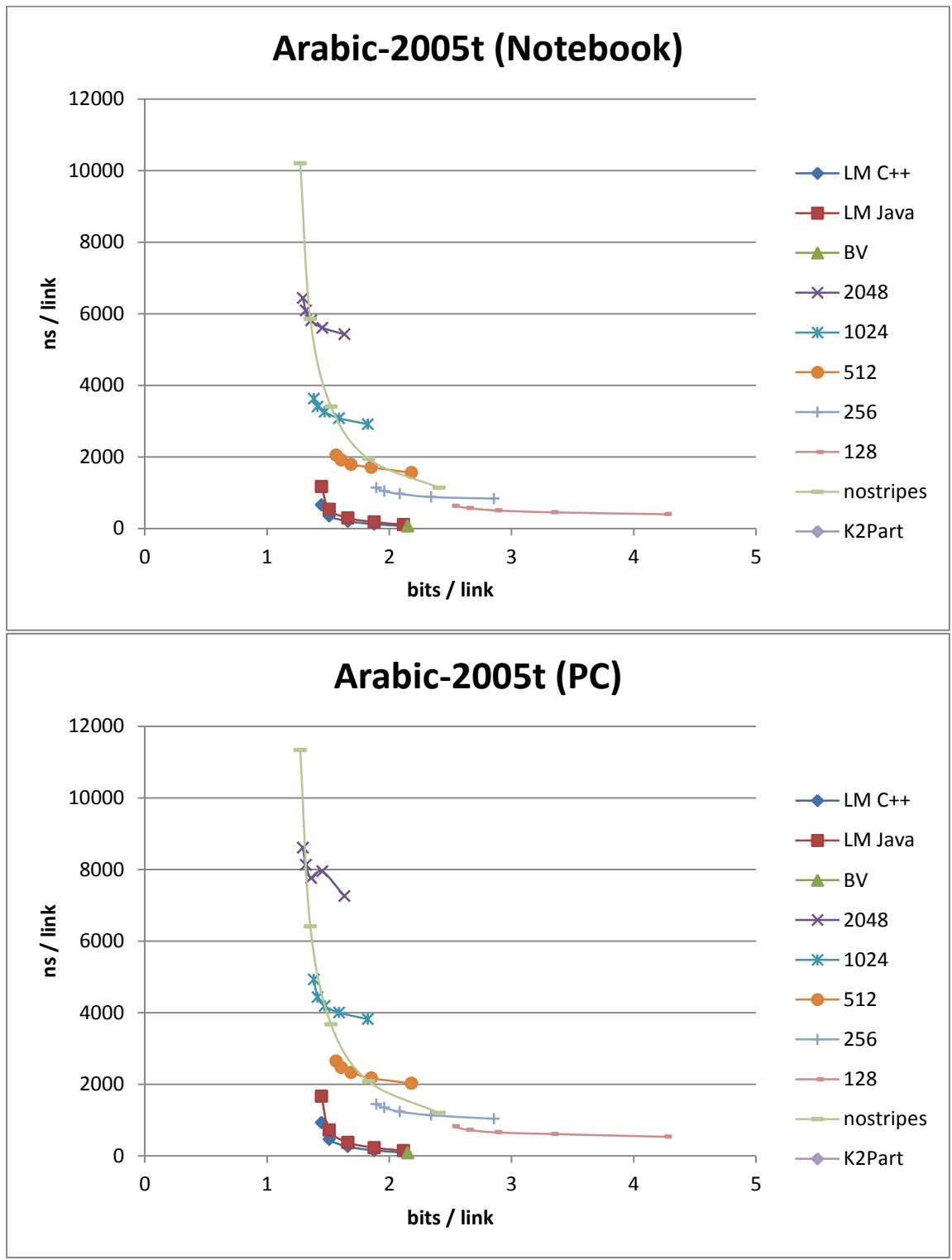

*Figure 16.Random access test diagrams. 2048, 1024, 512, 256 and 128 correspond to 2D stripes variants (B=2048, 1024, 512, 256 and 128 respectively). Five points constructing lines of 2D stripes are different values of K (K=8, 16, 32, 64 and 128 respectively).*

As the tests show LM method gives always more compact graph representations than BV does, whichever variant is chosen, and for LM-8 and LM-16 they are also faster in random access, however only in retrieving successor lists (LM method is slower than BV in transposed matrices). This can be explained by the fact, that BV copies the list from some other node chosen previously, which can contain similar number of successors while LM decodes the whole list of residuals and then checks whether it is a successor for the current node or not. Since the maximum number of residues in a transposed matrix is a couple of times larger (9 905 for Arabic-2005 and 575 618 for Arabic-2005t), the longer time is needed to process the whole query even when there is no successor at all.

Again, as with compression time, the C++ implementation of LM proved to be faster than its Java counterpart this time over 50%. This impressive result supports the statement that smartly written C++ is more efficient in processing data than Java.

Unfortunately 2D algorithms are still at least one order of magnitude behind 1D methods when it comes to random access time, however the difference has dropped thanks to 2D stripes variant. It decreased access time by 26.1% on average increasing size by 23%. All 2D stripes $K$ variants are faster in retrieving successor list than 2D non-striped algorithm yielding the same $B$ parameter.

$K^2$partitioned produced very similar output for all four different compression levels. Still provided bigger outcome than 2D nostripes and also slower random access time than $B = 128$ variant.

### 6.4. Discussion of the results

As may be observed all tests demonstrate similar behaviour characteristic for each test set. The results of random access times tests proved to be stable and reliable. The Indochina-2004 set showed interesting feature while being compressed with 2D stripes, B=2048 when the increasing $K$ made huge difference in access times. This may result from adjacency matrix being sparse and there are many tiles that do not have to be processed.

Most of the times 2D stripes algorithm produces too large output using $K \geq 64$. This is not proportional to the access time reduction, so practically it is inutile. On average access time was decreased in similar degree as the compression ratio increased comparing to 2D nostripes algorithm, so the 2D stripes method followed the tendency of its predecessor filling the gaps between consecutive tile sizes $B$.

BV in practice produced bigger output maintaining low random access times similar to those of LM. The only difference was noticed using transposed web graphs when the compression ratio was as good as in the fastest LM method, LM-8 (which produces the biggest outcome).

Another thing the test showed was usefulness of rewriting Java code into C++ as the performance gain was huge. However, in the field of web graphs, most people tend to use Java as its projects are easier to code and to integrate with other solutions.

$K^2$ partitioned on average produced the largest output out of all 2D methods. However, random access times were rather satisfactory with being beaten only by 2D (both striped and non-striped) with tile size $B = 128$.

## 6.5. Directions for future work

2D stripes algorithm has a great potential as the tests showed really impressive compression ratios among other methods. In order to improve access time it is necessary to increase decompression speed or find another method of representing the contents of a tile. Here the Re-Pair method could be implemented or simpler compressor may be used like LZ78 with external dictionary. Furthermore since most of the time there exist many tiles in a row, their decompression could be parallelized.

# 7. SUMMARY

The thesis presented satisfied all the assumptions formulated at the beginning and provided solid bases for future experiments on web graph compression. It has been proved that in this fast developing field there is still room for new achievements. New 2D variant is just one example of such possibilities. Furthermore if efficiency and low resource consumption is taken into consideration, the C++ programming language is definitely better choice than other popular languages like Java or Python.

# 8. REFERENCES

[1]  P. Boldi and S. Vigna, "The webgraph framework I: Compression techniques," *In Proc. of the Thirteenth International World Wide Web Conference (WWW 2004),* pp. 595-601, 2004.

[2]  F. Claude and S. Ladra, "Practical representations for web and social graphs," *Proceedings of the 20th ACM International Conference on Information and knowledge management CIKM '11,* pp. 1185-1190, 2011.

[3]  M. Sima, *Web Graph Visualization, Bachelor's Thesis,* Łódź, 2011.

[4]  P. Boldi, M. Santini and S. Vigna, "Permuting web and social graphs," *Internet Math, 6(3),* pp. 257-283, 2010.

[5]  P. Boldi, M. Rosa, M. Santini and S. Vigna, "Layered label propagation: A multiresolution coordinate-free ordering for compressing social networks," *In Sadagopan Srinivasan, Krithi Ramamritham, Arun Kumar, M. P. Ravindra, Elisa Bertino, and Ravi Kumar, editors, Proceedings of the 20th International Conference on World Wide Web,* pp. 587-596, ACM, 2011.

[6]  G. Buehrer and K. Chellapilla, "A scalable pattern mining approach to web graph compression with communitites," *Proceedings of the International Conference on Web search and web data mining WSDM '08,* pp. 95-106, 2008.

[7]  A. Apostolico and G. Drovandi, "Graph Compression by BFS," *Algorithms, 2(3) (ISSN 1999-4893),* pp. 1031-1044, 2009.

[8]  Y. Asano, Y. Miyawaki and T. Nishizeki, "Efficient compression of web graphs," *IEICE Transactions,* pp. 2454-2462, 2009.

[9]  G. Navarro and F. Claude, "A fast and compact Web graph representation," *Proceedings of the 14th International Conference on String processing and information retrieval SPIRE'07,* pp. 118-129, 2007.

[10]  V. N. Anh and A. Moffat, "Local modeling for webgraph compression," *2010 Data Compression Conference DCC,* p. 519, 2010.

[11]  N. R. Brisaboa, S. Ladra and G. Navarro, "K2-trees for compact web graph representation," *Proceedings of the 16th International Symposium on String Processing and Information Retrieval SPIRE '09,* pp. 18-30, 2009.

[12]  S. Grabowski and W. Bieniecki, "Tight and simple Web graph compression," *In: Holub,*

*J.,Žd'árek, J. (eds.) Proceeding of the Prague Stringology Conference,* pp. 127-137, 2010.

[13] S. Grabowski and W. Bieniecki, "Merging adjacency lists for efficient Web graph compression," *Man-Machine Interactions 2 AISC 103,* pp. 385-392, 2011.

[14] J. N. Larsson and A. Moffat, "Off-line dictionary-based compression," *Proceedings of the Data Compression Conference, DCC 1999, Snowbird, Utah, USA,* pp. 296-305, 1999.

# 9. ATTACHMENTS

## 9.1. Contents of a DVD

- Final thesis in a DOCX, DOC and PDF format.

- Visual Studio 2010 solution files (LM, 2D, 2D-stripes).

- BV test Java code and $k^2$ partitioned sources compiled for Win x64 platform.

- Zipped data sets used for testing.

- Used literature.

- Test logs and results (XLSX, XLSM format).

## 9.2. Figures

### 9.3. Code snippets

### 9.4. Tables

# 10.  GLOSSARY

RAM – Random Access Memory – main memory of a PC

BV – Boldi-Vigna standard for compressing web graphs

WWW – World Wide Web – the most popular service in the Internet

LM – List Merge: 1D method of compressing web graphs

$K^2$part – $k^2$partitioned: 2D method of compressing web graphs

LAW – Laboratory for Web Algorithmics: research lab of web and social networks

DSI – Computer Science Department at University of Milan

Indegree – the number of nodes directing at a given node

Outdegree – the number of nodes directed to by a given node

Web graph – structure representing hyperlinks between web documents

URL – Uniform Resource Locator: unique identifier of a web document

LLP – Layered Label Propagation: method of permuting URLs developed at LAW

VNM – Virtual Node Miner: technique of decreasing the number of edges

BFS – Breadth First Search: special type of traversing the graph

Interlink – the link that directs to the document in the same domain

Intralink – the link that directs to the document outside the domain

Re-Pair – technique of reducing edges by substituting pairs with new symbol

Deflate – compression algorithm based on Huffman coding and LZ77

LZ78 – improved version of LZ77 compression algorithm

STL – Standard Template Library: set of structures and containers for C++